\PassOptionsToPackage{
  pdfencoding=auto,
  psdextra,
  colorlinks=true,
  linkcolor=blue,
  citecolor=blue,
  urlcolor=blue
}{hyperref}

\documentclass[aps,prb,showpacs,floatfix,amsmath,amssymb,
superscriptaddress,reprint,nofootinbib]{revtex4-2}

\usepackage{amsthm}
\usepackage{float}
\usepackage{tikz}
\usepackage{subcaption}
\usetikzlibrary{arrows.meta}
\usepackage{bbold}
\usepackage[normalem]{ulem}
\newtheorem{proposition}{Proposition}
\newtheorem{corollary}{Corollary}
\usepackage{color}
\usepackage{graphicx}
\usepackage{mathrsfs}
\usepackage{epsfig}
\usepackage{setspace}
\usepackage{amsmath}
\usepackage{amssymb,amsfonts,mathrsfs,mathtools,braket}
\usepackage{verbatim}
\usepackage{bibentry}
\usepackage{caption}
\usepackage[T1]{fontenc}
\usepackage{pxfonts}
\usepackage{cancel}
\usepackage[percent]{overpic}
\usepackage{xcolor}

\usepackage{hyperref}
\usepackage{orcidlink}
\newcommand{\bea}{\begin{eqnarray}}
\newcommand{\eea}{\end{eqnarray}}

\begin{document}
\title{Krylov Complexity: Flat bands and Carroll breaking deformations}
\author{Aritra Banerjee\,\orcidlink{0000-0002-9619-7682}}\email{aritra.banerjee@pilani.bits-pilani.ac.in}
\affiliation{Birla Institute of Technology and Science, Pilani Campus, Rajasthan 333031, India}
\author{Arpan Bhattacharyya\,\orcidlink{0000-0002-7933-6441}}\email{abhattacharyya@iitgn.ac.in}
\affiliation{Department of Physics, Indian Institute of Technology Gandhinagar, Gujarat 382055, India}
\author{Rudranil Basu\,\orcidlink{0000-0003-0655-0890}}\email{rudranilb@goa.bits-pilani.ac.in}
\affiliation{Department of Physics, Birla Institute of Technology and Science Pilani, Zuarinagar, Goa 403726, India}
\author{Sayan Das\,\orcidlink{0009-0000-2420-767X}}\email{p20240080@pilani.bits-pilani.ac.in}
\affiliation{Birla Institute of Technology and Science, Pilani Campus, Rajasthan 333031, India}

\begin{abstract}
\medskip
   Systems with flat band structures, when written in the language of Compact Localised States (CLS), have been shown to be explicitly invariant under supertranslation symmetries, making Carrollian symmetries inherently important for such systems. In this work, we explore the state dynamics of these systems, focusing on quenches induced by Carroll breaking perturbations, through the probe of Krylov (spread) Complexity. We specialise to Fermionic ladder Hamiltonians with all bands flat (ABF) scenario, augmented by a supertranslation preserving interaction, and discuss Krylov state complexity for quenches across critical lines. We further discuss how the growth of Krylov complexity sharply resolves the phase-dependent resilience of Carrollian sectors against delocalising perturbations. This is augmented by a complementary mechanism for Krylov growth in a continuum Carroll scalar field theory with a gradient deformation, which exhibits strong ultraviolet sensitivity (UV/IR mixing). 
\end{abstract}
\maketitle
\section{Introduction}
Flat electronic bands have emerged as a central theme of interest in contemporary condensed matter physics \cite{Tovmasyan,thumin,zurita,danieli2020many,danieli2022many,mahyaeh2022superconducting,Aoki:2025drm}. Spanning systems as diverse as twisted bilayer graphene \cite{tarnopolsky2019origin} and moiré heterostructures \cite{cao2018correlated}, to engineered photonic and cold-atom lattices \cite{vicencio2015observation} , dispersionless bands provide a fertile platform for strong correlation effects \cite{peschel2003calculation,thumin}, unconventional superconductivity \cite{volovik2018graphite,aoki2020theoretical}, and topological order \cite{sachdev1999quantum,FQHE}. A ubiquitous feature of flat bands is the existence of compact localised states (CLS) \cite{leykam2018artificial,cls1,CLS2,CLS3}, which are single-particle eigenstates with strictly finite spatial support arising from destructive interference \footnote{For experimental observation of CLS in photonic lattices, see \cite{vicencio2015observation}.}. In contrast to Bloch waves, CLS do not propagate even in the presence of finite hopping amplitudes, leading to ultra-local dynamics and extensive degeneracy of single-particle eigenstates.

Despite being an interesting arena of research, flat bands have mostly been addressed at a phenomenological level, since a unifying principle to classify such scenarios has mostly been absent \footnote{See, however, efforts to classify flat bands using various algebraic methods, including  \cite{origami, xu2020building, danieli2024flat,morfonios2021flat, blockparti} }. 
Recently, it has been recognised that flat bands are not merely a consequence of fine-tuned lattice geometry but are deeply connected to emergent spacetime symmetries \cite{Bagchi:2022eui, Ara:2024fbr}. In particular, it was recently shown that flat-band lattice models \cite{yang2012topological} admit an infinite-dimensional algebra of conserved charges, identifiable with \textit{Carrollian supertranslations} \cite{Bondi:1962px, Sachs} \footnote{See \cite{Bagchi:2025vri} for a review on Carrollian physics. For more holography oriented introduction, see \cite{Ruzziconi:2026bix} as well.}, implying an emergent Carroll symmetry in the limit of vanishing effective propagation velocity \cite{Ara:2024fbr}. In general Carrollian theories appear when the characteristic velocity scales in the system vanish, most notably important for a famous Inönü-Wigner contraction of the Poincaré algebra where the speed of light goes to zero \cite{SG, LBLL, Henneaux:1979vn}. In this limit, the Hamiltonian densities at different spatial points commute with each other, and the system becomes ultra-local in the precise sense that time evolution is generated independently at each spatial site. This Carrollian viewpoint provides a unified framework for understanding CLS, flat dispersions \cite{aoki1996hofstadter,huber2010bose,ramachandran2017chiral}, and the absence of transport in a single symmetry based paradigm.

A particularly concrete realisation of these ideas is provided by the Creutz ladder \cite{Creutz:1999sd} type models at the special point where the two hopping amplitudes are equal \cite{leykam2018artificial}. At this point, the spectrum becomes completely flat, and the Hamiltonian can be written in terms of CLS modes ($\alpha_j,\beta_j$), engineered as combination of site local fermions, that are strictly localised on plaquettes. Moreover, \cite{Ara:2024fbr}  also demonstrated that one can introduce interactions constructed solely from these CLS modes in a manner that preserves the supertranslation symmetry, resulting in an exactly solvable yet nontrivial interacting many-body system. Remarkably, this interacting Carroll-symmetric model, even in the ultra-local setting, exhibits multiple quantum phases at half filling, including conventional/\textit{Vanilla} phases with unique CLS product ground states and an \textit{Exotic} phase with a macroscopically degenerate ground state manifold not spanned by CLS alone. This latter phase can be shown to host very interesting quantum phenomena stemming from a plethora of states defined by different symmetrisation.

A central result in \cite{Ara:2024fbr} is that these distinct phases respond in sharply different ways to Carroll-breaking perturbations. Introducing a small asymmetry between the Creutz ladder hoppings breaks the explicit supertranslation invariance at the flat-band point which restores finite band dispersion and generates a nonzero effective Fermi velocity. In the interacting theory, this perturbation induces long-range correlations and delocalisation as the critical line is approached from the vanilla phase, while the exotic phase remains ultra-local and displays a  robustness of its correlation functions. In this sense, the exotic phase retains finite remnants of Carrollian protection even away from the exact symmetry points. This sharp dichotomy strongly suggests that Carroll symmetry imprints phase-dependent structural constraints even on the interacting spectrum, with the exotic phase inheriting an emergent resilience that stabilises localised quantum behavior beyond the strict flat-band limit.

However, correlation functions and Loschmidt echoes, as discussed above, probe only limited aspects of many-body dynamics. They do not directly characterise how extensively a quantum state explores the Hilbert space under time evolution, nor whether the system develops ergodic operator growth \cite{Serbyn:2020wys,turner2018weak}. A recently developed and useful diagnostic of such behaviour is \textit{Krylov complexity} \cite{parker2019universal} or Spread Complexity \cite{balasubramanian2022quantum} when specialising to states rather than operators \footnote{The readers are directed to some excellent current reviews \cite{sanchez2024krylov,nandy2025quantum,rabinovici2025krylov,baiguera2026quantum} (and references therein) on this topic for a more well-rounded introduction.}. Given a Hamiltonian $H$ and an initial state $|\psi\rangle$, the Krylov construction builds an orthonormal basis by repeated action of $H$, leading to a tridiagonal (Lanczos) representation \cite{lanczos1950iteration,viswanath1994recursion}. This form is characterised by Lanczos coefficients $b_n$ which encode the connectivity of successive Krylov sectors and serve as dynamical probes. The time-dependent spread of $|\psi(t)\rangle$ in this Krylov basis defines a complexity $C_K(t)$ that quantifies operator growth, delocalisation in Hilbert space, and the onset of chaos or thermalisation.

Krylov complexity has proven to be a sharp probe of quantum chaos \cite{rabinovici2022krylov,rabinovici2022krylov2,hashimoto2023krylov,erdmenger2023universal,Bhattacharyya:2023grv,espanol2023assessing,camargo2024spread,scialchi2024integrability,alishahiha2025krylov,baggioli2025krylov,balasubramanian2025quantum,baggioli2025quantum}, integrability and thermalisation \cite{Bhattacharjee:2022qjw,nandy2024quantum,alishahiha2025thermalization,caputa2025complexity}, phase transition \cite{Caputa_2022,Banerjee:2022ime,Bhattacharya:2023yec,Beetar:2023mfn,bento2024krylov,anegawa2024krylov,Takahashi_2025,Chakrabarti:2025hsb,jiang2026krylov} and many-body localisation \cite{ballar2022krylov,menzler2024krylov}. Further, it has been connected to operator spreading and hydrodynamic behaviour in both lattice systems \cite{rabinovici2022krylov, rabinovici2022krylov2,Fan:2022xaa,Bhattacharya:2023zqt, Aguilar-Gutierrez:2023nyk,murugan2026schwinger,bhattacharyya2026stochastic,grabarits2026} and quantum field theories \cite{dymarsky2021krylov,avdoshkin2024krylov,camargo2023krylov,he2024probing,vasli2024krylov}\footnote{These references are in no way exhaustive, the interested reader is directed to various references associated to these works.}. Yet, its behaviour in flat band systems with emergent Carroll symmetry (and breaking thereof) remains essentially unexplored \footnote{However, for Conformal Field Theories, the flow to Carrollian cousins thereof were probed using geometric complexity in \cite{Banerjee:2022ime}.}. In particular, it is not known how CLS-based ultra-locality and the exotic Carroll phases discovered in \cite{Ara:2024fbr} would manifest themselves in Krylov space once supertranslation symmetry is weakly broken by a relevant operator (see the discussion in \cite{banerjee2023one} for relevant operators in Carroll field theories). Especially, from a spread complexity point of view, this is a pertinent question, as the interacting flat band system supports a diverse spectrum of different physically interesting states, for which the complexity growth under perturbations would be distinct, especially at early times. More precisely, we would like to understand the spread complexity signature of supertranslation breaking in such a system. 

In this work, we address this question directly by systematically investigating spread (Krylov) complexity in our interacting model subjected to controlled Carroll-breaking perturbations. We discuss the mechanism of spread of states in the interacting, supertranslation invariant model to start with. Then we focus on quench protocols that start from physically distinguished ground states of the Carroll-symmetric theory, viz:
(i) the CLS-product ground state of the Vanilla phase and (ii) representative ground states of the Exotic phase, including domain-wall and translation-symmetrised configurations. We show that these two classes of states display radically different Krylov dynamics. Vanilla phases are extremely brittle and spread quickly as soon as the perturbation is turned on.
  More interestingly, since the exotic phase is highly degenerate, differently symmetrised initial states are differently susceptible to spreading in the Krylov space. Some exotic phase states exhibit rapid Krylov growth and quickly explore a much larger fraction of the many-body Hilbert space than others. Some particular classes, on the other hand, remain frozen and show no discernible spread. We then devise an invariant to quantify this differential growth of states, at least for early time, living in the same degenerate manifold. We also elucidate on the late time behaviour of Krylov growth across parameter space.

Our results establish Krylov complexity as a powerful dynamical tool of Carroll-symmetry protection/destruction in flat band systems. They provide a quantitative, basis-independent measure of the fragility of vanilla phases and the high initial state dependence of the degenerate vacuum manifold in exotic Carroll phases, thereby extending the static and correlation function–based analyses to the full many-body Hilbert space. We further comment on the universality of this quantifier in generic Carroll breaking scenarios, going even beyond lattice theories, since such a large degeneracy of states is a generic hallmark of ultra-locality. For this purpose, we also discuss a scalar field theory with inherent Carroll symmetry, leading to ultra-local correlations. Adding a Carroll relevant derivative term to this theory implies growth in Krylov space. We show how this spread leads to Lanczos coefficients being explicitly dependent on ultraviolet scales, another Carrollian characteristic with deep theoretical significance.

The rest of the paper is organised in the following way. Section \eqref{secII} introduces our setup and the phase structure of the interacting supertranslation invariant theory. After introducing spread complexity and other Krylov space probes in Section \eqref{secIII}, we formalise the single particle spectrum and introduce the quench protocols for the perturbed theory. Further, Section \eqref{secIV} contains the main results of the paper, that quantify the early time complexity growth for different initial states, via the introduction of active link invariants for the relevant states. In section \eqref{secV} we offer a unique perspective on Krylov growth in Carroll systems by considering a ultra-local scalar field deformed by a spatial gradient. We end our discussion with some interesting outlook in Section \eqref{secend}. Appendices contain some more details of our computation and various additional discussions. 

\section{Model and Carroll-symmetric phases}\label{secII}

Let us begin by introducing our setup, where the hosted all flat-bands setup can be explained using an emergent supertranslation symmetry. The basics of the model has already been introduced in \cite{Ara:2024fbr}, where the authors concentrated on a first principle construction of the compact localised states as a direct consequence of the supertranslation symmetry on the lattice. In what follows, we will suppress some details, and only focus on the main aspects important for the current discussion. 

\subsection{Creutz ladder and compact localised states}

To elucidate our ideas, we consider the Creutz ladder, a two-leg fermionic lattice model with Hamiltonian
\begin{equation}
H_0=\sum_{j}\Big[
t_1\big(c^\dagger_{j+1}c_j-d^\dagger_{j+1}d_j\big)
+t_2\big(c^\dagger_{j+1}d_j-d^\dagger_{j+1}c_j\big)
+\text{h.c.}
\Big],
\label{eq:creutz}
\end{equation}
where $c_j,d_j$ are fermionic operators on the two legs of the ladder and $t_{1,2}$ are intra and inter-chain hopping parameters.  In momentum space, the single-particle Hamiltonian is
\begin{equation}
H(k)=2t_1\cos k\,\sigma_3-2t_2\sin k\,\sigma_2 .
\end{equation}
Once we diagonalise the kernel, it gives the two energy bands
\begin{equation}
E_\pm(k)=\pm\sqrt{2(t_1^2+t_2^2)+2(t_1^2-t_2^2)\cos(2k)} .
\label{preflat}
\end{equation}

At the special point $t_1=t_2\equiv \tau$, the dispersion becomes completely flat,
$E_\pm=\pm 2|\tau|$, and the model admits a basis of compact localised states (CLS) that survive without any effective propgation amongst them.
Defining the CLS operators out of the site local fermions:
\begin{equation}
\alpha_j =
\frac{1}{2}
\left(
c_j+d_j+c_{j+1}-d_{j+1}
\right),
\quad
\beta_j =
\frac{1}{2}
\left(
c_j+d_j-c_{j+1}+d_{j+1}
\right).
\label{eq:CLS}
\end{equation}
the Hamiltonian \eqref{eq:creutz} reduces to the ultra-local form
\begin{equation}
H_0=2\tau\sum_j\big(\alpha_j^\dagger\alpha_j-\beta_j^\dagger\beta_j\big).
\label{eq:ultralocal}
\end{equation}
In this representation, the Hamiltonian is diagonal in real space, and the CLS basis as written with $\alpha_j,\beta_j$
are strictly localised on individual plaquettes. The time dependent correlation function of $\alpha$ or $\beta$ modes are exactly ultra-local; for example, given $\tau>0$, the ground state of the model is the one where all $\beta$ modes are filled (Vanilla-$\beta$ state), and two-point functions simply read:
\begin{equation}
    \langle \beta_i^\dagger(t)\beta_j(0)\rangle
=
e^{-2 i \tau t}\,\delta_{ij},~~~
\langle \alpha_i^\dagger(t)\beta_j(0)\rangle = 0
=
\langle \alpha_i^\dagger(t)\alpha_j(0)\rangle .
\end{equation}

At the gapless limit, this is nothing but the discrete avatar of the spatial delta function one gets for continuum Carroll Conformal fermions \cite{Ara:2024fbr, Bagchi:2022eui, Banerjee:2022ocj, Hao:2022xhq, Bergshoeff:2023vfd}.
The commutativity of the Hamiltonian densities
at different lattice sites, i.e.:
\begin{equation}\label{DSC}
    [H_i,H_j] = 0~~ \forall~~(i,j)
\end{equation}
implies an infinite set of conserved charges \footnote{For a lattice setting this would technically be the number of lattice points.}, which can be identified with
Carrollian supertranslations \footnote{Note that the vanishing of Dirac-Schwinger condition \cite{Henneaux:2021yzg} for the continuum Carroll field theories dictates that Hamiltonian densities at two spatial points commute with each other, giving a tell-tale sign of the emergent symmetry. This is just a discrete lattice version of the same.}. These supertranslations, which work as angle dependendent translations in spacetime \cite{Bondi:1962px, Sachs}, act very particularly on CLS states defined on lattice points:
\begin{equation}
    \delta_f \xi_j = f_j\dot{\xi}_j,
\end{equation}

where $f_j$ are arbitrary functions defined on individual sites and the dot represents derivative with time. 
As a result, the model at $t_1=t_2$ realises an interacting lattice
version of a Carroll-symmetric ultra-local quantum theory. This theory, as explained,  naturally hosts an all bands flat scenario.

\subsection{Supertranslation-invariant interactions}

Following \cite{Ara:2024fbr}, one may add interactions built from purely CLS operators to our ladder model, that preserve
supertranslation symmetry and hence ultra-locality.  The simplest such interaction is the on-site
density--density coupling
\begin{equation}
H_{\rm int}=V\sum_j n_j^\alpha n_j^\beta ,
\end{equation}
where the number operators $n_j^{\alpha,\beta}=\alpha_j^\dagger\alpha_j,\beta_j^\dagger\beta_j$.
The full interacting, but still Carroll-symmetric, Hamiltonian is therefore
\begin{equation}
H_{\rm C}=\sum_j\Big[V n_j^\alpha n_j^\beta+2\tau(n_j^\alpha-n_j^\beta)\Big].
\label{eq:HC}
\end{equation}
Even with the interaction added, one can explicitly check that the Hamiltonian on two lattice sites commute, i.e. \eqref{DSC} still holds, implying the model is still exactly solvable and possesses an
extensively degenerate spectrum.

Throughout the paper, unless stated otherwise, $L$ denotes the number of physical \textit{rungs} of the ladder. Each rung carries two CLS orbitals, $\alpha_j$ and $\beta_j$, so the single-particle Hilbert space has $2L$ modes. Half filling means total fermion number $N=L$. For even $L$, the deformation $H_\Delta$ preserves the two sublattices of rung parity, and each parity sector contains $L/2$ rungs, or equivalently $L$ CLS modes. In this sector the model \eqref{eq:HC} exhibits a rich phase
structure despite its ultra-locality. This phase diagram, spanned by the parameters $V$ and $\tau$, is shown in Figure  \eqref{phaseplot}. In this phase diagram we still have the all-$\beta$ and all-$\alpha$ filled ground states as in the non-interacting model, but it is supplemented by a new phase consisting of highly degenerate configurations.

\subsection{Vanilla and Exotic Carroll phases}
\begin{figure}
    \centering
    \includegraphics[width=0.95\linewidth]{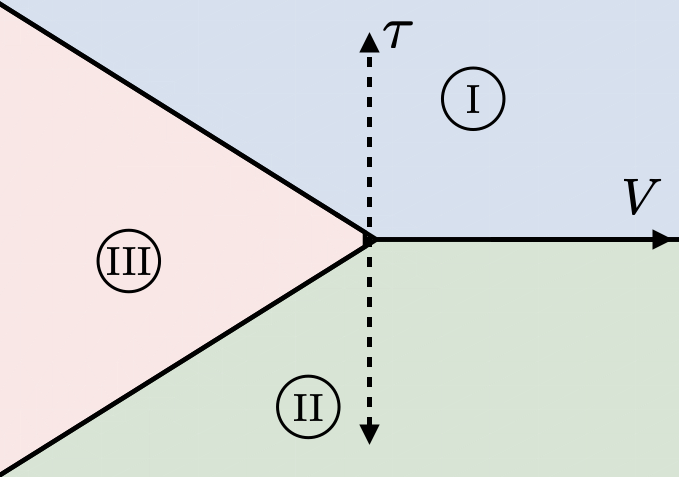}
    \caption{The phase plot corresponding to the Hamiltonian \eqref{eq:HC}. Regions I, II, and III are, respectively, the Vanilla-$\beta$, Vanilla-$\alpha$, and Exotic phases. The discussion in the main text focuses on the \(\tau>0\) side, where the Vanilla-\(\beta\) phase meets the Exotic phase at \(V=-4\tau\).}
    \label{phaseplot}
\end{figure}

Referring to Figure \eqref{phaseplot},
for the vanilla phase defined by $\tau>0$ and $V>-4\tau$, the unique half-filled ground state of \eqref{eq:HC} is the
CLS product state
\begin{equation}
|\Psi_{\beta}\rangle=\prod_j \beta_j^\dagger |0\rangle ,
\label{eq:betaGS}
\end{equation}
which we referred earlier to as the Vanilla-$\beta$ phase.  Similarly, for $\tau<0$ and
$V>4\tau$ the ground state is the corresponding all-$\alpha$ product state
(Vanilla-$\alpha$ phase).  In both cases, the ground state is a simple CLS product with
zero correlation length and area-law entanglement.
In contrast, for $V<0$ and $|4\tau|<|V|$, the half-filled ground state becomes macroscopically
degenerate.  A convenient basis for this \emph{Exotic} Carroll phase consists of all CLS
configurations satisfying
\begin{equation}
n_j^\alpha=n_j^\beta, \, ~ \forall j,\, ~~
\sum_j n_j^\alpha=\sum_j n_j^\beta=\frac{L}{2}.
\label{eq:exotic}
\end{equation}
With the convention fixed above, $L$ is the number of rungs and $2L$ is the number of CLS orbitals, equivalently the number of original ladder sites. Equation~\eqref{eq:exotic} therefore says that the exotic ground-state basis contains $L/2$ doubly filled rungs (\textit{Doublon}) and $L/2$ \textit{empty} rungs. States in this phase have zero overlap with all $\alpha$ and all $\beta$ filled states. The dimension of this degenerate ground state is 
$\binom{L}{L/2}$. Curiously, on the critical line $V = -4\tau$, the degeneracy is (see Appendix \eqref{statecount} for details):
\begin{equation}
    D_0 = \sum_{\mathfrak{g}=0}^{\lfloor L/2 \rfloor} \frac{L!}{\mathfrak{g}! \cdot \mathfrak{g}! \cdot (L-2\mathfrak{g})!}
\end{equation}
which grows exponentially with system size $\sim 3^L$.  While the Hamiltonian remains ultra-local through all the phases, the degeneracy of the ground state leads to interesting physics. Curiously, the supertranslation invariant states in this case are not lattice translation covariant. This leads one to consider different physically motivated choices for the ground state in this regime, including the possibility of translation symmetrised domain-wall and other extended many-body states.

A key result of \cite{Ara:2024fbr}, as mentioned previously, is that these phases respond in sharply different ways to Carroll-breaking perturbations. In the vanilla phases, any perturbation that breaks supertranslation invariance induces long-range correlations and rapid delocalisation, whereas in the exotic phase, ultra-locality of correlation functions remains robust, in at least a class of well behaved states from the degenerate ground state manifold. In the following section, we will show that this dichotomy has an even more nuanced fine-structured manifestation in the growth of Krylov complexity, which is the main focus of the current work.

\section{Carroll-breaking deformation and Krylov construction}\label{secIII}

\subsection{Krylov construction}
Since the interacting model introduced in the last section still has an infinite dimensional supertranslation symmetry, we would now like to consider the setup where such a symmetry is broken. To quantify our probe,
we now briefly review the Krylov construction for a many-body Hamiltonian, following
\cite{rabinovici2022krylov}.  Given a Hamiltonian $H$ and an
initial normalised state $|\psi_0\rangle$, one constructs an orthonormal Krylov basis
$\{|r_n\rangle\}$ by repeated action of $H$:
\begin{equation}
|r_{n+1}\rangle=\frac{1}{b_{n+1}}\Big(H|r_n\rangle-a_n|r_n\rangle-b_n|r_{n-1}\rangle\Big),
\end{equation}
with $|r_0\rangle=|\psi_0\rangle$ and $b_0=0$.  The coefficients
\begin{equation}
a_n=\langle r_n|H|r_n\rangle,\qquad
b_{n+1}=\big\|H|r_n\rangle-a_n|r_n\rangle-b_n|r_{n-1}\rangle\big\|
\end{equation}
define the Lanczos tridiagonalisation of $H$ in the Krylov basis \cite{viswanath1994recursion}.

We now would like to ask, how complex a given initial state becomes upon evolution over arbitrarily length of times?
The time-evolved state $|\psi(t)\rangle=e^{-iHt}|\psi_0\rangle$ may then be expanded in a convenient basis
\begin{equation}
|\psi(t)\rangle=\sum_{n=0}^{K-1}\phi_n(t)|r_n\rangle ,
\end{equation}
where $K$ is the dimension of the Krylov space. In fact one can think of the system in Krylov subspace gets mapped to that of a single particle hopping on a semi-infinite tight-binding chain defined by the Lanczos coeffiecients.  The Krylov complexity is defined as the first moment of the probability distribution of finding this fictitious particle at a site on the tight binding model:
\begin{equation}\label{CK_def}
C_K(t)=\sum_{n=0}^{K-1} n\,|\phi_n(t)|^2 ,
\end{equation}
which measures how far the state has spread from the initial vector in Krylov space. A closely related quantity is the Krylov Inverse Participation Ratio (KIPR) written in the same eigenbasis,
\begin{equation}
\mathrm{KIPR}(t)=\sum_{n=0}^{K-1}|\phi_n(t)|^4 ,
\end{equation}
which probes localisation in Krylov space. This ratio measures how many states in the preferred basis has effectively participated over time evolution of the system. These two quantities are of primary interest in the following analysis.
\subsection{Example: Single particle Krylov spectrum}

To understand the Lanczos tridiagonalisation for the single particle sector, let us start by considering the non-interacting model away from flat-band, i.e. when $t_1\neq t_2$ in \eqref{eq:creutz}. The hamiltonian in this case reads:
\begin{equation}
\begin{aligned}
H =
\sum_j \Big[
& (t_1+t_2)\,\alpha_j^\dagger \alpha_j
-(t_1+t_2)\,\beta_j^\dagger \beta_j
\\
&+\frac{t_1-t_2}{2}
\left(
\alpha_{j+2}^\dagger \alpha_j
+
\alpha_j^\dagger \alpha_{j+2}
\right)
+
\frac{t_2-t_1}{2}
\left(
\beta_{j+2}^\dagger \beta_j
+
\beta_j^\dagger \beta_{j+2}
\right)
\\
&+\frac{t_1-t_2}{2}
\left(
\beta_{j+2}^\dagger \alpha_j
-
\alpha_{j+2}^\dagger \beta_j
+
\alpha_j^\dagger \beta_{j+2}
-
\beta_j^\dagger \alpha_{j+2}
\right)
\Big].
\end{aligned}
\label{eq:H_CLS_general}
\end{equation}
Starting with the one $\alpha$ filled initial state $|r_0\rangle = |0, \alpha\rangle$, the Lanczos coefficients take very simple form:
\begin{align}
& a_0 = t_1+t_2,\qquad a_n=0\quad(n\geq 1),
\nonumber
\\
&b_{2k-1}=|t_1-t_2|,\qquad
b_{2k}=|t_1+t_2|\quad(k\geq 1).
\label{eq:single_particle_lanczos}
\end{align}
having, quite expectedly for a period-2 dimerised chain, only alternating hopping amplitudes and on-site energy terms.

To probe further, we could investigate how wavepackets spread in the single particle case, i.e. find an estimation for the group velocity. For this, one uses Bloch wavefunctions at the lattice sites, and because the period is two, a two‑site unit cell is used. Let the on-site wavefunctions be
\begin{equation}
\psi_{2K}   = A\,e^{i 2K k},~~
\psi_{2K+1} = B\,e^{i(2K+1)k},~~ k\in[-\pi/2,\pi/2].
\end{equation}
Putting these back into the tight-binding equations give us: 
\begin{align}
E A &= (b_1 e^{ik} + b_2 e^{-ik}) B, \nonumber \\
E B &= (b_1 e^{-ik} + b_2 e^{ik}) A.
\end{align}
Eliminating $(A,B)$ from the above gives the dispersion relation
\begin{equation}
%E^2(k) = b_1^2 + b_2^2 + 2b_1 b_2 \cos(2k),
%\qquad
E(k) = \pm\sqrt{b_1^2 + b_2^2 + 2b_1 b_2 \cos(2k)},
\end{equation}
which is nothing but the spectrum \eqref{preflat} written in terms of the single particle Lanczos coefficients. 
Give this, The group velocity, for the positive branch, which dominates the wave packet, is
\begin{equation}
v_g(k) = \left|\frac{dE}{dk}\right|
        = \frac{2b_1 b_2\,|\sin(2k)|}{\sqrt{b_1^2 + b_2^2 + 2b_1 b_2 \cos(2k)}}.
\end{equation}
This clearly vanishes when $t_1=t_2$, i.e. no propagation at the flat-band point. Also, the maximum group velocity can be found after a bit of algebra:
\begin{equation}
v_{g,\max} = 2\,\min(b_1,b_2) = 2\,\min\!\big(|t_1-t_2|,\,|t_1+t_2|\big).
\end{equation}
At least at the single particle level, one can then compare this situation with spacetime light-cones, which close down at Carroll limits. In this case, the characteristic velocity is given by the group velocity which similarly vanishes at the flat-band point. This single-particle calculation only motivates the velocity scale. In the interacting $(V\neq 0)$ half-filled problem, the relevant propagating object is the particle-hole excitation which will be analysed below. We will, however, return to the group velocity discussion in a later section. 

\subsection{Adding Carroll-breaking perturbation}

To probe the stability of the Carroll-symmetric phases described in the earlier section, we introduce
a controlled breaking of supertranslation invariance by detuning the hopping amplitudes of
our \textit{interacting} Creutz ladder model \eqref{eq:HC},
\begin{equation}
t_1=\tau+\Delta,\qquad t_2=\tau-\Delta ,
\end{equation}
with $|\Delta|\ll |\tau|$ a perturbation parameter around the flat band point.  In the original site basis, this simply corresponds to moving away
from the supertranslation invariant point.  When expressed in the CLS basis \eqref{eq:CLS}, however, this deformation induces nonlocal hopping between CLS modes.  The full Hamiltonian becomes
\begin{equation} \label{Htot}
H=H_{\rm C}+H_\Delta ,
\end{equation}
where $H_{\rm C}$ is the ultra-local interacting Hamiltonian \eqref{eq:HC} and
\begin{align}
H_\Delta=\Delta\sum_j \Big(
&\alpha^\dagger_{j+1}\alpha_{j-1}-\beta^\dagger_{j+1}\beta_{j-1}
+\beta^\dagger_{j+1}\alpha_{j-1}
\nonumber\\
& -\alpha^\dagger_{j+1}\beta_{j-1} + \text{h.c.}\Big).
\label{eq:Hdelta}
\end{align}
This term explicitly breaks supertranslation invariance and ultra-locality, allowing
excitations in the CLS basis to propagate through the lattice.

As discussed previously, going away from flat-band point restores dispersive dynamics. In fact, in the free theory ($V=0$), this deformation generates a finite Fermi velocity. To see that, one can expand the dispersion relation written in terms of Lanczos coefficients near the band minima and maxima. Near the valence band maxima this is a parabolic (Schrödinger) band with an effective mass:
\begin{equation}
 m^*_{\max} = \frac{1}{\frac{d^2{E}}{dk^2}}|_{k=0}
       = \frac{b_1 + b_2}{4b_1 b_2}=\frac{1}{8}\left(\frac{1}{\Delta} + \frac{1}{\tau}\right) ,
\end{equation}
ignoring higher order terms.
Similarly for the same near band minima, 
we can set \(k = \frac{\pi}{2} + q\) with \(|q|\ll 1\), so that we get a Poincaré like dispersion relation $E(q) = -\sqrt{\Delta_{\min}^2 + v_F^2 q^2}$, where the rest energy (gap) and effective Fermi velocity are given by:
\begin{align}
     \Delta_{\min} &= |b_1 - b_2| ,\nonumber \\
     v_F &= 2\sqrt{b_1 b_2} = 4\sqrt{\tau\Delta}.
\end{align}
Note that in both cases, putting $\Delta=0$ gives us the ultra-localised flat-band situation, but in one case it is achieved by making the vacuum infinitely massive, in the other case the ground state fermi velocity vanishes. 

  We thus see, in the Carroll breaking scenario, $H_\Delta$ is the
only source of spatial correlations and transport, and therefore provides a natural
control parameter for studying the fate of Carroll-symmetric phases. When we turn on interactions ($V\neq 0$), all of this becomes extremely sensitive to initial states and the phase they belong to.

\subsection{Real-space propagation of the soft half-filled exciton}
\label{subsec:soft-exciton-spreading}

The discussion above shows that the deformation $H_\Delta$ in \eqref{eq:Hdelta}
is the microscopic source of spatial propagation away from the Carroll point.  Before turning
to Krylov growth in the full many-body Hilbert space, it is useful to isolate the elementary
low-energy object whose motion is responsible for the spreading of correlations near the
half-filled transition when interactions are also included.  This gives a direct real-space diagnostic of the same Carroll-breaking
mechanism with $V\neq 0$.

To do that, we focus on the Vanilla-$\beta$ reference state introduced in \eqref{eq:betaGS}, i.e. $|\Psi_\beta\rangle=\prod_{j=1}^{L}\beta_j^\dagger |0\rangle$,
and consider one particle-hole excitation (i.e. a doublon-empty pair) above it,
\begin{equation}
    |m,n\rangle
    =
    \alpha_m^\dagger \beta_n |\Psi_\beta\rangle .
\end{equation}
Here $m$ labels the position of the $\alpha$ particle and $n$ labels the position of the
$\beta$ hole.  The important point is that the same-site and off-site excitons have different
energies under the ultra-local Hamiltonian $H_{\rm C}$ in \eqref{eq:HC}.  Measured relative
to the all-$\beta$ state,
\begin{equation}
    E_{m=n}-E_{\beta}=4\tau ,
    \qquad
    E_{m\neq n}-E_{\beta}=4\tau+V .
    \label{eq:soft-exciton-gap}
\end{equation}
Thus the excitation that softens near the half-filled transition is not the local same-site
flip $\alpha_j^\dagger\beta_j|\Psi_\beta\rangle$, but the off-site exciton
$\alpha_m^\dagger\beta_n|\Psi_\beta\rangle$ with $m\neq n$.  The corresponding soft gap is
\begin{equation}
    \epsilon_{\rm ex}=4\tau+V .
\end{equation}
This explains why the correlation matrix becomes spatially extended close to
$V=-4\tau$: the low-energy particle-hole excitation itself becomes soft.

To make this mechanism explicit, we project the Hamiltonian
\begin{equation}
    H=H_{\rm C}+H_\Delta
\end{equation}
onto the one-exciton subspace
\begin{equation}
    \mathcal H_{\rm ex}
    =
    {\rm span}\left\{
    |m,n\rangle=\alpha_m^\dagger\beta_n|\Psi_\beta\rangle
    \right\}.
\end{equation}
Let $P_{\rm ex}$ be the projector onto this subspace.  The projected Hamiltonian is
\begin{equation}
    H_{\rm ex}=P_{\rm ex}(H_{\rm C}+H_\Delta)P_{\rm ex}.
\end{equation}
Using \eqref{eq:Hdelta}, one obtains
\begin{align}
    H_{\rm ex}
    &=
    \sum_{m,n}
    \varepsilon_{mn}
    |m,n\rangle\langle m,n|
    \nonumber\\
    &\quad
    +2\Delta
    \sum_{m,n}
    \left(
    |m+2,n\rangle\langle m,n|
    +
    |m-2,n\rangle\langle m,n|
    \right)
    \nonumber\\
    &\quad
    -2\Delta
    \sum_{m,n}
    \left(
    |m,n+2\rangle\langle m,n|
    +
    |m,n-2\rangle\langle m,n|
    \right),
    \label{eq:projected-exciton-H}
\end{align}
with periodic boundary conditions and
\begin{equation}
    \varepsilon_{mn}
    =
    4\tau+V(1-\delta_{mn}) .
\end{equation}
Remember, away from the contact line $m=n$, the $\alpha$ particle and $\beta$ hole hop independently.
The first hopping term in \eqref{eq:projected-exciton-H} propagates the
$\alpha$ particle by two rungs, while the second propagates the $\beta$ hole by two rungs
with the opposite sign.  Considering the two‑member wavefunction as a product of plane waves,
$\ket{q_\alpha,q_h} \sim \sum_{m,n} e^{i(q_\alpha m + q_h n)}\ket{m,n}$, the off-site exciton therefore
has the approximate two-body dispersion
\begin{equation}
    E(q_\alpha,q_h)
    =
    \epsilon_{\rm ex}
    +
    4\Delta\cos(2q_\alpha)
    -
    4\Delta\cos(2q_h).
    \label{eq:exciton-dispersion}
\end{equation}
The corresponding group velocities are then given by $\frac{\partial E}{\partial q}$:
\begin{equation}
    v_\alpha(q_\alpha)
    =
    -8\Delta\sin(2q_\alpha),
    \qquad
    v_h(q_h)
    =
    8\Delta\sin(2q_h),
\end{equation}
so the characteristic propagation speed is set by
\begin{equation}
    v_{\rm ex}^{\rm max}\sim 8|\Delta|.
\end{equation}
Thus $H_\Delta$ turns the soft near-critical exciton into a mobile object with speeds set by $|\Delta|$.  This is the
real-space counterpart of the correlation-matrix spreading discussed in the Carroll-broken
half-filled problem \cite{Ara:2024fbr}. So all in all, even in the interacting case $H_\Delta$ is the sole agent that turns localised states into propagating modes.

For the numerical demonstration of this exciton spreading, we initialise a localised antisymmetric exciton packet at the
center rung $j_0$,
\begin{equation}
    |\psi_{\rm ex}(0)\rangle
    =
    \frac{1}{\sqrt{2}}
    \left(
    |j_0+1,j_0-1\rangle
    -
    |j_0-1,j_0+1\rangle
    \right).
    \label{eq:initial-soft-exciton}
\end{equation}
Writing
\begin{equation}
    |\psi_{\rm ex}(t)\rangle
    =
    e^{-iH_{\rm ex}t}|\psi_{\rm ex}(0)\rangle
    =
    \sum_{m,n}\psi_{m,n}(t)|m,n\rangle ,
\end{equation}
we monitor the $\alpha$-particle and $\beta$-hole distributions
\begin{align}
    p_\alpha(j,t)
    &=
    \sum_n |\psi_{j,n}(t)|^2 ,
    \\
    p_h(j,t)
    &=
    \sum_m |\psi_{m,j}(t)|^2 .
\end{align}
The spreading width for the exciton is defined using the shortest distance $d(j,j_0)$ between rung $j$ and the central rung on the ring,
\begin{align}
    R_\alpha^2(t)
    &=
    \sum_j d(j,j_0)^2 p_\alpha(j,t),
    \\
    R_h^2(t)
    &=
    \sum_j d(j,j_0)^2 p_h(j,t),
    \\
    R^2(t)
    &=
    \frac{1}{2}
    \left[
    R_\alpha^2(t)+R_h^2(t)
    \right].
    \label{eq:exciton-R2}
\end{align}

\begin{figure}
    \centering
    \begin{subfigure}[htb]{0.48\textwidth}
        \centering
        \includegraphics[width=\linewidth]{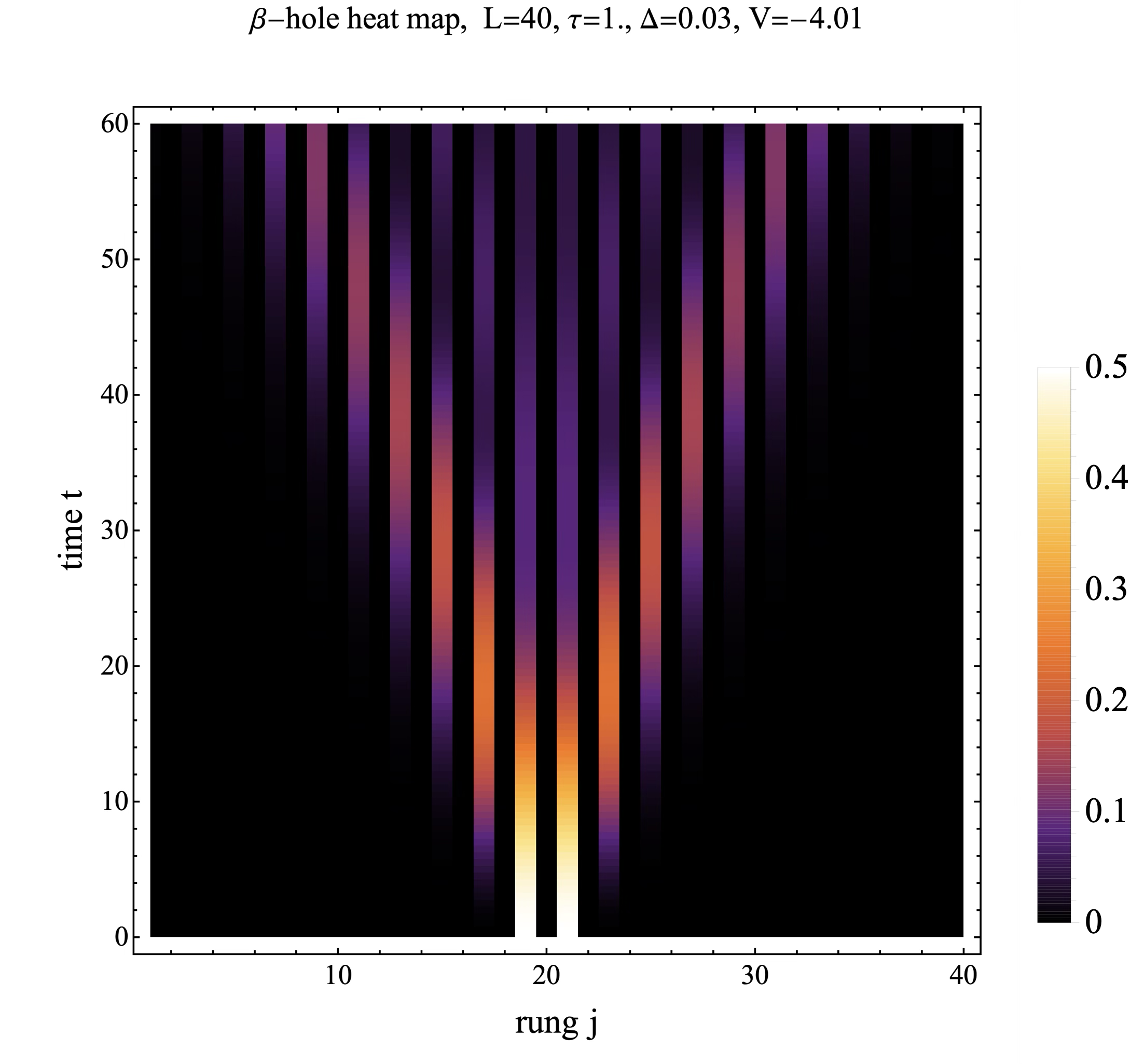}
        \caption{
        $\beta$-hole density $p_h(j,t)$ for the localised soft-exciton packet in
        \eqref{eq:initial-soft-exciton}.
        }
        \label{fig:soft-exciton-hole}
    \end{subfigure}
    \hfill
    \begin{subfigure}[htb]{0.48\textwidth}
        \centering
        \includegraphics[width=\linewidth]{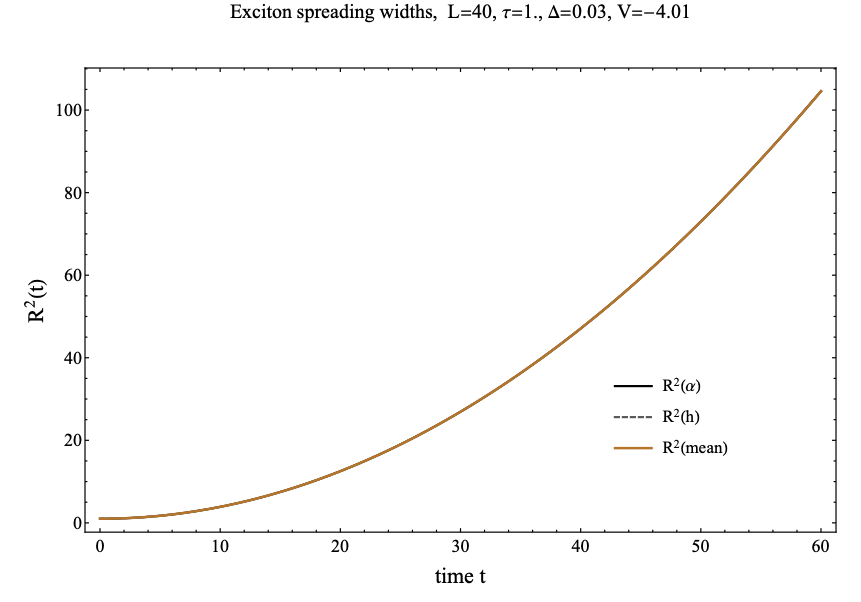}
        \caption{
        Exciton spreading width $R^2(t)$ defined in \eqref{eq:exciton-R2}.
        The approximately quadratic growth is consistent with ballistic propagation.
        }
        \label{fig:soft-exciton-R2}
    \end{subfigure}
    \caption{
    Real-space propagation of the soft half-filled exciton near the Carroll critical line.
    The calculation is performed in the projected one-exciton subspace
$|m,n\rangle=\alpha_m^\dagger\beta_n|\Psi_\beta\rangle$ for
    $L=40$, $\tau=1$, $\Delta=0.03$, and $V=-4.01$.
    The soft off-site exciton has energy scale $4\tau+V$, while the deformation
    $H_\Delta$ supplies a finite propagation velocity.  The heat map in panel (a)
    shows the ballistic spreading of the $\beta$ hole, and panel (b) shows the
    corresponding growth of the second moment.  This provides a real-space diagnostic
    of the same Carroll-breaking mechanism that later drives Krylov-space spreading.
    }
    \label{fig:soft-exciton-spreading}
\end{figure}

Figure \eqref{fig:soft-exciton-spreading} shows the resulting propagation for
$L=40$, $\tau=1$, $\Delta=0.03$, and $V=-4.01$, so that the system is very close to the
soft-exciton line $V=-4\tau$.  The projected one-exciton Hilbert space has dimension
$L^2=1600$.  The $\beta$-hole density develops a clear fan-shaped (or light-cone shaped) profile, while the second
moment grows approximately quadratically before finite-size effects become important.  This
is the expected signature of ballistic spreading generated by the Carroll-breaking
deformation.  In this sense, Figure \eqref{fig:soft-exciton-spreading} gives a real-space
visualisation of the soft mode whose many-body consequences will be probed below through
Krylov complexity.

\subsection{Quench protocols}\label{sec:quenches}
As of now, we have been trying to quantify how excitations propagate in our system when the Carroll breaking perturbation is turned on. 
Moving on to full numerics for an exact picture, we now consider quench protocols in which some initial state $|\psi_0\rangle$ is an exact
ground state of the Carroll-symmetric full interacting Hamiltonian $H_{\rm C}$ \eqref{eq:HC}, while the quenching Hamiltonian includes the Carroll-breaking perturbation $H_\Delta$ defined in \eqref{Htot}. The interaction strength $V$ is chosen such that the initial state belongs to a well-defined Carroll phase, which can be either of the following situations.

\paragraph{Vanilla quench:}
In the first protocol, the initial state is the unique half-filled ground state of the
Vanilla-$\beta$ phase mentioned before,
\begin{equation}
|\psi_0^{(1)}\rangle = \prod_{j=1}^{L} \beta_j^\dagger |0\rangle ,
\end{equation}
which is a compact localised state (CLS) product, with a very small amount of entanglement, with respect to the real space degrees of freedom, i.e. the site local modes $c_i, d_i$. For completeness, this state can also be represented in the rung (orbital) form:
\begin{equation}\label{rungb}
  |Vanilla_{\beta}\rangle=  \Bigg| 
  \begin{array}{c}
    \medcirc\medcirc \dots\medcirc\medcirc\\
    \medbullet \medbullet\dots\medbullet\medbullet
  \end{array} \Bigg\rangle , ~~ \
\end{equation}
where the filled circles correspond to occupied levels.
The system is then evolved with the Hamiltonian
\begin{equation}
H = H_{\rm C} + H_\Delta ,
\end{equation}
For the Vanilla-reference quench we evolve this state close to the critical line from the adjacent side,
\begin{equation}
V=-4\tau-\epsilon,\qquad 0<\epsilon\ll\tau,\qquad |\Delta|\ll\tau .
\end{equation}

This choice keeps the quench resonant with the soft off-site exciton scale $|4\tau+V|=\epsilon$ while testing how rapidly the CLS product state loses Carroll protection when the symmetry-breaking perturbation is switched on.

\paragraph{Exotic phase quench:}
In the second protocol, the initial state is chosen from the macroscopically degenerate ground-state manifold of the Exotic Carroll phase, of which there could be many classes:
\begin{equation}
  |\psi_0^{(2)}\rangle =  \left(\prod_{i=1}^{N_1\leq L} \alpha^\dagger_{m_i}\right)
\left(\prod_{j=1}^{N_2\leq L} \beta^\dagger_{n_j}\right)
|0\rangle
\end{equation}
Concretely, we can consider the example of a domain-wall configuration in the CLS basis, consisting of consecutive regions of
$\alpha$- and $\beta$-occupied rungs at half filling:
\begin{equation}
 |DW\rangle = \prod_{j=1}^{L/2} \alpha_j^{\dagger}\beta_j^\dagger |0\rangle ,
\end{equation}
%This state is not a CLS product and represents a typical element of the Exotic ground-state manifold. 
but this is only a typical element, represented by:
\begin{equation}\label{rungdw}
    |DW\rangle = \Bigg| 
  \begin{array}{c}
     \overbrace{\medbullet \medbullet\dots \medbullet\medbullet}^{L/2}\medcirc\medcirc\dots\medcirc\medcirc\\
    \underbrace{\medbullet \medbullet\dots \medbullet\medbullet}_{L/2}\medcirc\medcirc\dots\medcirc\medcirc
  \end{array} \Bigg\rangle.
\end{equation}
and the set is much larger.

For these second kind of quenches, the initial vectors are chosen from the ground-state manifold of the exotic side and are evolved with the same Carroll-breaking Hamiltonian $H_\Delta$ near the line from the vanilla side,
\begin{equation}
V=-4\tau+\epsilon,\qquad 0<\epsilon\ll\tau,\qquad |\Delta|\ll\tau .
\end{equation} 
The two families therefore probe the same near-critical window from complementary phase-referenced initial data. In the exotic family, the degeneracy makes the choice of initial vector physically consequential, and the quench response is controlled by the local occupancy pattern of the selected state.

As should be clear, on a highly degenerate manifold the choice of actual initial state will determine how the quench acts on the system.

These two protocols contrast a unique CLS product state with a macroscopically degenerate set of doublon/empty product states under the same symmetry-breaking mechanism. The question we ask is therefore: \emph{how do phase-referenced Carroll states with very different local structure spread in Krylov space once $H_\Delta$ is switched on?}
\section{Krylov Dynamics in the perturbed model}\label{secIV}
In this section we study the growth of Krylov complexity following Carroll-breaking
quenches starting from distinct ground states of the Carroll-symmetric interacting Hamiltonian
as introduced before. All Krylov constructions are performed in exact invariant blocks of the half-filled Hilbert space. We always impose total particle-number conservation. In addition, because $H_\Delta$ connects only rungs of the same parity, the odd- and even-rung particle numbers $N_o$ and $N_e$ are separately conserved. The numerical calculations below are therefore carried out in the fixed-particle, rung-parity-resolved block specified in Appendix~\eqref{numerical_recipe}. This is an exact symmetry reduction of the lattice Hamiltonian, not an energetic or low-energy projection.
\subsection{Susceptibility for states: Active links}
Before we dig into the \textit{vulnerability} or \textit{robustness} of the ground state(s) of these Vanilla/Exotic phases through Krylov growth using full numerical machinery, it is useful to understand the action of $H_{\Delta}$ on the states.  
This can be done through noting two separate operators, arising through a $\mathbb{Z}_2$ grading of the lattice, i.e. by defining the odd and even operators:
\begin{equation}
    N_o = \sum_{j \, \mathrm{odd}} (n^{\alpha}_j + n^{\beta}_j), \, N_e = \sum_{j \, \mathrm{even}} (n^{\alpha}_j + n^{\beta}_j). 
\end{equation}
Since $H_{\Delta}$ contains \textit{active bonds} between either odd-indexed rungs or even-indexed ones,
\bea \label{Z2}
[H_{\Delta}, N_o ] =0= [H_{\Delta},N_e]
\eea
follows trivially.

Let us now restrict ourselves, specifically to the degenerate ground state manifold of the $V<-4\tau$ (Exotic) phase. Recalling that these states consists of rungs, which are either completely filled or completely empty \eqref{eq:exotic}, let us introduce the binary to pinpoint the occupancy number
\begin{equation} \label{dj}
     d_{j} = \left\{ \begin{array}{ll}
 1, & \text{if rung } j \text{ is filled}, \\
 0, & \text{if rung } j \text{ is empty}.
 \end{array} \right.
\end{equation}
Hence, the global constraint is $\sum_j d_j = L/2$, equivalent to \eqref{eq:exotic}. This doublon/empty structure can be better understood looking at the rung representation of states drawn in the previous section (\eqref{rungdw}). 

Let the normalised initial state \(|\psi_0\rangle\) belong to the doublon/empty manifold, then representative states can be written as:
\begin{equation}
|\psi_0\rangle = |\{d_j\}\rangle,
\qquad
d_j\in\{0,1\},
\qquad
\sum_j d_j=\frac{L}{2},
\end{equation}
where, as explained before, \(d_j=1\) denotes a doublon
$|D\rangle_j=\alpha_j^\dagger\beta_j^\dagger|0\rangle_j,
$
and \(d_j=0\) denotes an empty site. Since, active bonds in $H_{\Delta}$ (see \eqref{eq:Hdelta}) are only between $j-1$ and $j+1$ rungs, the state represented by $|\{d_j\}\rangle$ in this ground state manifold can be usefully characterised by the following degree of active bonds, defined as:
\begin{eqnarray}
    A(\{d_j\})&:= & \sum_{j=1}^L d_{j-1} (1-d_{j+1}) +  d_{j+1} (1-d_{j-1}) \nonumber\\
    & =&\sum_{j=1}^L (d_{j-1} - d_{j+1})^2
\end{eqnarray}
Let's call $(d_{j-1} - d_{j+1})^2$ the \textit{Active Link} number $\chi_j$ for this case.
To see how this is important, note that by definition, the Lanczos coefficients
\begin{equation}
b_1=\|(H-a_0)|\psi_0\rangle\|,
\qquad
a_0=\langle\psi_0|H|\psi_0\rangle.
\end{equation}
Since \(|\psi_0\rangle\) in our case is a doublon/empty product state, the onsite piece \(H_C\) is diagonal on it:
\begin{equation}
H_C|\psi_0\rangle = E_0 |\psi_0\rangle,
\end{equation}
with \(E_0=(L/2)V\) at half filling. Moreover,
\begin{equation}
\langle\psi_0|H_\Delta|\psi_0\rangle=0,
\end{equation}
because \(H_\Delta\) changes the local occupation pattern of the initial state, it therefore has no diagonal matrix element on \(|\psi_0\rangle\). Thus
\begin{equation}
a_0=E_0,
\qquad
(H-a_0)|\psi_0\rangle = H_\Delta|\psi_0\rangle,
\end{equation}
where $H_\Delta|\psi_0\rangle$ goes out of the doublon/empty manifold. Consequently we also have
\begin{equation}
b_1^2=\|H_\Delta|\psi_0\rangle\|^2.
\end{equation}
Now write
\begin{equation}
H_\Delta = 2\Delta \sum_j T_j,
\end{equation}
where the operator
\begin{equation}
T_j=
\alpha^\dagger_{j+1}\alpha_{j-1}
-\beta^\dagger_{j+1}\beta_{j-1}
+\beta^\dagger_{j+1}\alpha_{j-1}
-\alpha^\dagger_{j+1}\beta_{j-1}
+\mathrm{h.c.}
\end{equation}
acts only on the same-parity pair \((j-1,j+1)\). If the pair \((j-1,j+1)\) is inactive, i.e.\ \(D\!-\!D\) or \(0\!-\!0\), then
\begin{equation}
T_j|\psi_0\rangle=0,
\end{equation}
because hopping is either Pauli blocked or there is no particle to move. Hence \(\|T_j|\psi_0\rangle\|^2=0\).

If the pair \((j-1,j+1)\) is active, i.e. has a \ \(D\!-\!0\) or \(0\!-\!D\), then \(T_j\) produces exactly \textit{four} distinct two-site Fock states by moving one fermion from the doublon site to the empty site. See Appendix \eqref{Prop_App} for details of this counting. These four states are mutually orthogonal and each appears with unit amplitude up to a sign. Therefore 
\begin{equation}
\|T_j|\psi_0\rangle\|^2=4
\qquad\text{iff}\qquad
\chi_j=1.
\end{equation}
Thus, in all cases involving doublon/empty fock seeds,
\begin{equation}
\|T_j|\psi_0\rangle\|^2 = 4\chi_j.
\end{equation}

Furthermore, for \(j\neq k\), the states \(T_j|\psi_0\rangle\) and \(T_k|\psi_0\rangle\) are orthogonal Fock states, so all cross terms vanish:
\begin{equation}
\langle\psi_0|T_jT_k|\psi_0\rangle=0,
\qquad j\neq k.
\end{equation}
Hence
\begin{align}
\|H_\Delta|\psi_0\rangle\|^2
&=
(2\Delta)^2 \sum_j \|T_j|\psi_0\rangle\|^2
\\
&=
(2\Delta)^2 \sum_j 4\chi_j
=
16\Delta^2 A.
\end{align}
Therefore we can read off the Lanczos coefficient:
\begin{equation}
b_1^2=16\Delta^2 A,
\qquad
b_1=4|\Delta|\sqrt{A},
\end{equation}
Note now, the short-time Krylov complexity obeys the simple growth law
\begin{equation}
K(t)\sim b_1^2 t^2 = 16\Delta^2 A\, t^2,
\qquad t\to 0.
\end{equation}
In particular, states with \(A= \sum_j \chi_j = 0\) are frozen at first order, whereas larger active-link count produces larger initial curvature in \(K(t)\).

Thus this active link count given by \(A\) provides an exact leading-order ordering principle within the highly degenerate manifold for exotic phase ground states: states with smaller \(A\) are more robust against the Carroll-breaking perturbation, while states with larger \(A\) are more dynamically fragile and exhibit faster initial complexity growth.

The all-$\beta$ vanilla reference state is included in the numerical comparison for a different, but equally controlled, reason. We reiterate, it is not a member of the doublon/empty exotic manifold, so the binary variable $d_j$ and the active-link invariant $A$ do not apply to it. Nevertheless, it is an eigenstate of $H_{\rm C}$ at $\Delta=0$ and the first Lanczos coefficient can be obtained by direct action of $H_\Delta$. For $|\Psi_\beta\rangle=\prod_{j=1}^{L}\beta_j^\dagger|0\rangle$, the only non-vanishing terms in $T_j$ are the cross-flavour moves that convert a filled $\beta$ orbital into an empty $\alpha$ orbital two rungs away. Explicitly,
\begin{equation}
T_j|\Psi_\beta\rangle=
\left(\alpha_{j-1}^\dagger\beta_{j+1}-\alpha_{j+1}^\dagger\beta_{j-1}\right)|\Psi_\beta\rangle,
\end{equation}
up to the conventional fermionic signs fixed by the ordering of modes. These $2L$ one-exciton states are mutually orthogonal on the ring, and hence
\begin{equation}
b_{1,\beta}^2=\|H_\Delta|\Psi_\beta\rangle\|^2=8L\Delta^2.
\label{eq:beta-b1}
\end{equation}
The all-$\beta$ curve in the figures should therefore be read as the Vanilla CLS-product benchmark, while the active-link formula $b_1^2=16\Delta^2 A$ organises the doublon/empty exotic family. The comparison is meaningful precisely because both curvatures are computed from the same Lanczos definition but from different local structures of the initial state.

\subsection{Differential growth structure}

So far we demonstrated our Carroll‑breaking perturbation is only able to delocalise excitations if the initial state contains at least one active bond, i.e. a local configuration where a doublon and an empty rung sit on the same parity sublattice and thus can exchange fermions.

To exemplify how states in the doublon/empty manifold grow under perturbation, let us consider a state, in which alternate rungs are filled, i.e. for which $d_j = (1+ (-1)^j)/2$, for $j=1,2, \dots, L$ - a charged density wave, $|\mathrm{CDW}_1\rangle $. In this state, the degree of active bonds, $A=0$ and hence, this is a null state for $H_{\Delta}$. This state would remain frozen under a quench by $H$ of \eqref{Htot}. However, $|\mathrm{CDW}_1\rangle $ is not invariant under translation. Rather, if we define $|\mathrm{CDW}_2\rangle $ by $d_j = (1- (-1)^j)/2$, for $j=1,2, \dots, L$ and take a combination $|\mathrm{CDW}_+ \rangle  = \frac{1}{\sqrt{2}} \left( |\mathrm{CDW}_1\rangle + |\mathrm{CDW}_2\rangle \right)$, that will become a translationally invariant state, as well as remain frozen, when quenched by the dynamics of the perturbed $H$.

At the opposite extreme, consider the period-four pattern
\begin{equation}
\ket{\mathrm{P4}}=\ket{1\,1\,0\,0\,1\,1\,0\,0\,\cdots}
\end{equation}
here the 1s and the 0s are the values of $d_j$ as per the definition \eqref{dj}.
This state maximizes the number of active same-parity links ($A=L$) and therefore exhibits the largest initial Krylov growth.
    \begin{figure}[H]
        \centering
        \includegraphics[width=0.98\linewidth]{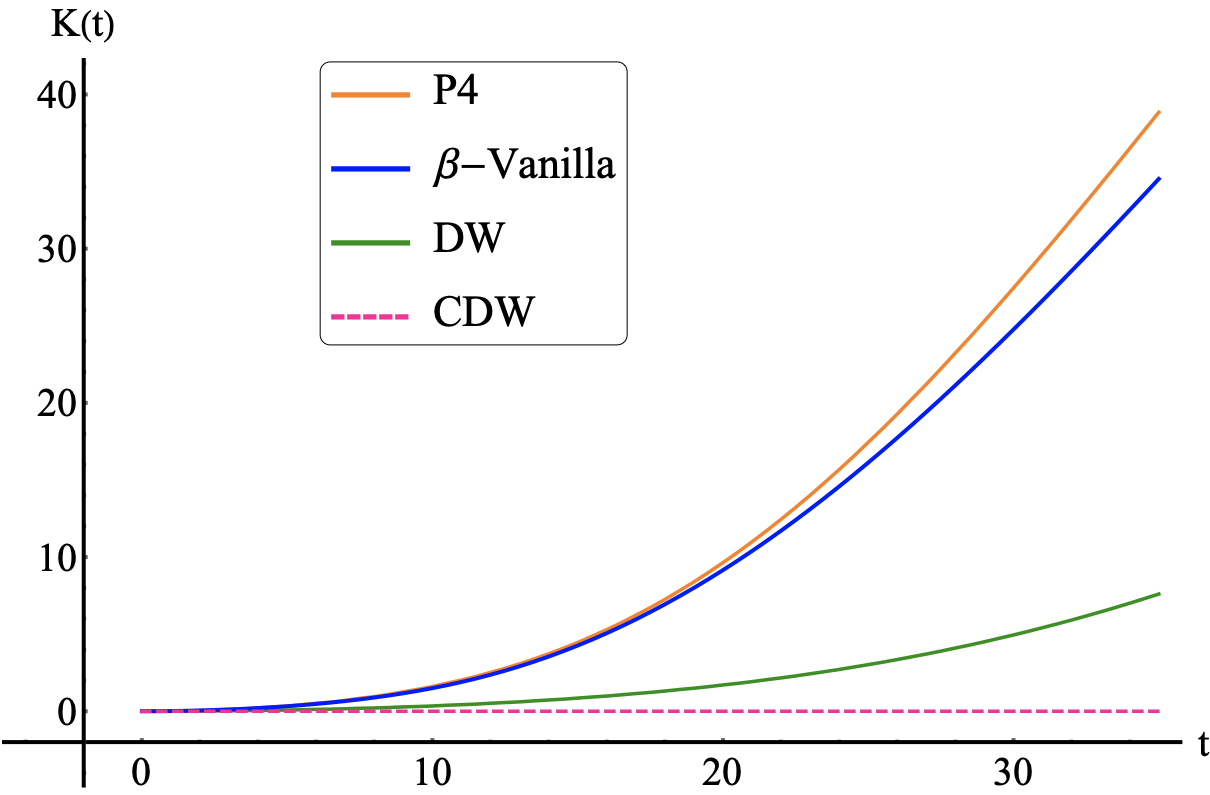}
        \caption{Krylov complexity growth for $L=16$ rungs, i.e. $32$ ladder sites. $\Delta = 0.01$. For the Vanilla-reference all-$\beta$ quench we use $V=-4\tau-\epsilon$, while for the Exotic-reference doublon/empty quenches we use $V=-4\tau+\epsilon$. We chose $\tau =1$ and $\epsilon = 10^{-4}$. The Lanczos calculation is performed in the exact fixed-$(N_o,N_e)$ parity block described in Appendix~\ref{numerical_recipe}.}
        \label{Krylov}
    \end{figure}

\begin{figure}[H]
    \centering
\includegraphics[width=0.98\linewidth]{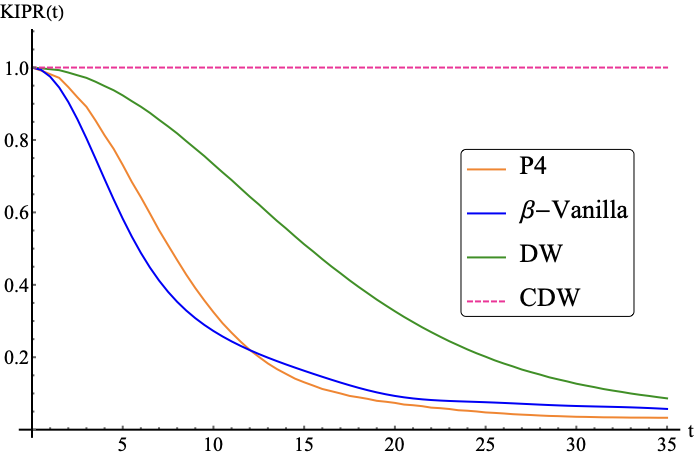}
    \caption{Early time KIPR for the four representative states. Other physical parameters are same as Figure  \eqref{Krylov}. For the Vanilla quench, the KIPR saturates at a comparatively lower value,
indicating strong spreading in Krylov space.  By contrast, the P4 state saturation value goes even lower, consistent with the discussion we have had in the text. DW states as usual take a middle path.}
    \label{kipr}
\end{figure}
    
Between these two limits lies the macroscopic domain wall,
\begin{equation}
\ket{\mathrm{DW}}=\ket{1\,1\,\cdots 1\,0\,0\,\cdots 0},
\end{equation}
for which \(A\) is nonzero but minimal among non-frozen configurations (for periodic boundary conditions, \(A=4\)).  In this sense, \(A\) defines a hierarchy of dynamical vulnerability within the \(\Delta=0\) ground-state manifold. The early-time growth of Krylov complexity and the KIPR has been plotted in Figure \eqref{Krylov} and \eqref{kipr} for the three above-mentioned doublon/empty states, and the $\beta-$ filled vanilla state under quench. The hierarchy of these sectors with different dynamical activity is very clear from these numerics. As should be clear already from previous discussions, the P4 exotic state actually grows faster than the highly brittle all-$\beta$ state! 

We also compare the hierarchy of Krylov growths of these states at a fixed time and varying values of the parameters $\Delta$ and $\epsilon$ in the Figure \eqref{heatmaprobust}. To compare the response of different initial states across the \((\epsilon,\Delta)\) plane, we evaluate the Krylov complexity at a fixed probe time \(t_\star\).  This time must be chosen within the common monotonic-growth window of all parameter points included in the scan; otherwise the resulting heat map would mix early-time growth with finite-size saturation and recurrence effects.  We therefore choose \(t_\star\) by estimating the fastest possible initial growth in the scanned region.

With our convention, $L$ is the total number of rungs and each parity sector contains $L/2$ rungs.  The fastest doublon/empty product state is the P4 pattern, for which the active-link count is maximal,
\begin{equation}
A_{\rm max}=L.
\end{equation}
Using the short-time result $K(t)\sim b_1^2t^2,~~
b_1^2=16\Delta^2 A$,
the largest initial growth rate in the scan is bounded by
\begin{equation}
b_{1,\rm max}^2
=
16\Delta_{\rm max}^2 A_{\rm max}
=
16L\,\Delta_{\rm max}^2 .
\end{equation}
\begin{widetext}
   \begin{minipage}{\linewidth}
\begin{figure}[H]
\centering
    \includegraphics[width=0.95\linewidth]{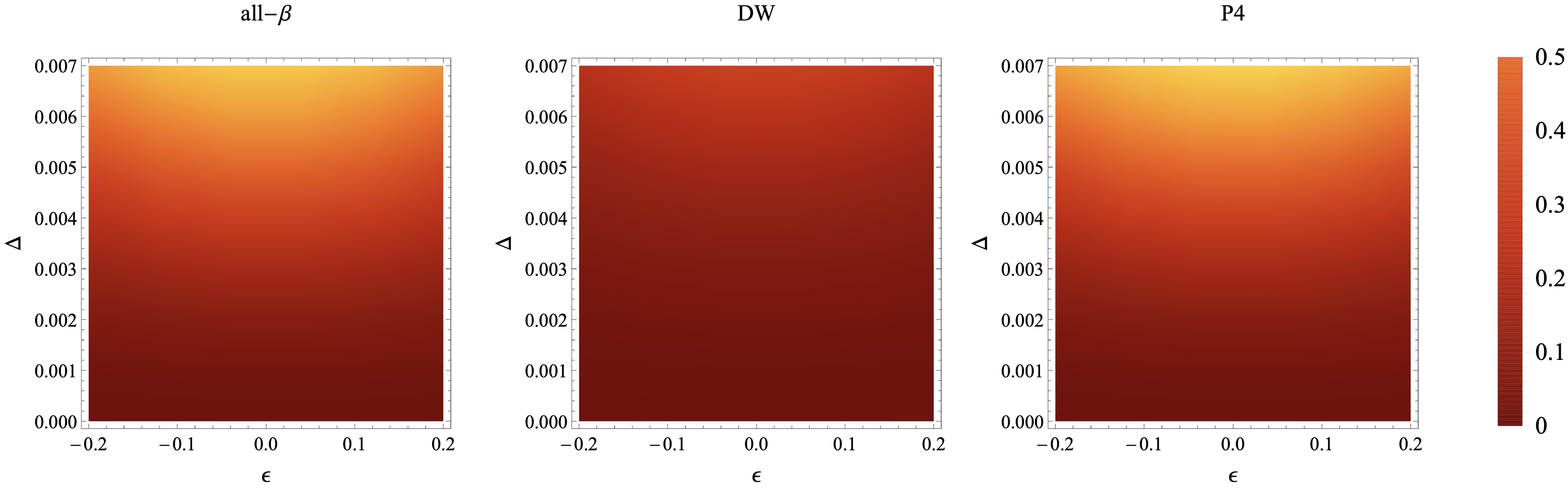}
    \caption{
    %A heatmap of early time complexity growth across parameter space. The growth is shown for all three states with non-zero growth as discussed in the text for different values of $\Delta$ and $\epsilon$. Simulation has been done for 12 rungs, at constant timeslice $t=10$. Lighter colors signify higher value of Krylov spreading. \textcolor{blue}{
    Finite-time Krylov response in the \((\epsilon,\Delta)\) plane at \(t_\star=10\), with \(V=-4\tau+\epsilon\) and \(\tau=1\).  The three panels show \(K(t_\star;\epsilon,\Delta)\) for the all-\(\beta\)-filled state, the doublon domain-wall state, and the P4 doublon pattern, respectively, using a common color scale.  The CDW state is not shown because it is an exact frozen state for all \((\epsilon,\Delta)\), with \(K_{\rm CDW}(t)=0\).  The maps display the expected hierarchy of dynamical vulnerability: the P4 and all-\(\beta\) states show the strongest response, while the domain wall remains much less susceptible to the \(\Delta\)-quench.}
    \label{heatmaprobust}
\end{figure}

\end{minipage}
\end{widetext}
A conservative common probe time is therefore
\begin{equation}
t_\star \lesssim \frac{\gamma}{b_{1,\rm max}}
=
\frac{\gamma}{4\Delta_{\rm max}\sqrt{L}},
\end{equation}
with \(\gamma=\mathcal O(1)\).  This choice ensures that even the most rapidly spreading state, at the largest \(\Delta\) in the scan, remains in the early monotonic regime.  The resulting heat maps should therefore be interpreted as finite-time early-growth response maps. They show how the leading active-link hierarchy is modified by finite-time detuning effects as one moves away from the resonant line \(V=-4\tau\).

%We stress that \(A\) is a sharp \emph{short-time} classifier: it fixes the initial curvature of \(K(t)\) exactly.  States with the same \(A\) may nevertheless differ at later times because of higher-order processes and, where relevant, different conserved odd/even particle-number sectors.  Nevertheless, \(A\) provides the natural organizing principle for comparing the quench dynamics of different ground states before turning to the full numerical time evolution.

\subsection{Physical meaning for Carroll systems}

%As we have seen, in Carroll-symmetric flat-band systems, the ultra-local Hamiltonian $H_{\rm C}$ acts independently on each CLS, implying that the Krylov chain generated from a CLS product state is strictly finite.  Consequently, the corresponding Krylov complexity is bounded and exhibits no growth.  By contrast, the Carroll-breaking term $H_\Delta$ generates hopping in the CLS basis and can induce extensive growth in Krylov space. The rate and extent of this growth depend sensitively on the nature of the initial state.
 
We should now pause for a bit and deliberate on what this active-link dependent initial growth means for Carroll invariance. The active-link structure provides an exact parametrisation of the initial Krylov vector in the low‑energy sector (responsible for the early time quadratic growth), and by tuning it one can explore a wide variety of dynamical trajectories — all eventually converging to the same ballistic growth, but with distinct transient behaviour that can be of physical interest. Most intriguingly, in this case, global Carroll invariance is broken, but local subsets of the Hilbert space (like the CDW sector) retain an effective Carroll symmetry due to their inertness. 
The `memory' of the flat phase is preserved in the active‑link hierarchy. Especially with zero links, one can be tempted to call the CDW states as \emph{scars} of the Carroll symmetry, however here they only arise due to a kinetic constraint, and they are not isolated special states \footnote{More explicitly, unlike quantum scars—which are exact eigenstates that avoid thermalisation—the CDW state is not an eigenstate of the full Hamiltonian}.

We should also talk about the structure of matrix elements for the Carroll breaking perturbation, which is further intriguing.
If $\ket{\psi}$ and $\ket{\phi}$ are pure doublon/empty product states,  i.e.  each rung $j$ is either
$\ket{0}_j \equiv \ket{n_j^\alpha = n_j^\beta = 0}$ or $\ket{D}_j \equiv \alpha_j^\dagger \beta_j^\dagger \ket{0}_j$,
then all matrix elements of $H_\Delta$ between them vanish:
\begin{equation}
\bra{\psi} H_\Delta \ket{\phi} = 0 \qquad \text{for any } \ket{\psi}, \ket{\phi} \in \big\{ \ket{0}, \ket{D} \big\}^{\otimes L}.
\end{equation}
This happens because a single action of $H_\Delta$ changes the occupation patterns. Note that this means first‑order degenerate perturbation theory within the ground‑state manifold gives exactly zero. This robustness under deformations has already been discussed in \cite{Ara:2024fbr}. However, The first Krylov step in this case leaps out of the doublon/empty manifold, and the chain grows in the full Hilbert space via ``virtual'' single–single states.
This strong selection rule at first order is very much specific for the Carroll case.

\subsection{Late time behaviour}
 
At strictly early times, the hierarchy between the different initial states, controlled by the active-link count, is quite sharp: the CDW state is frozen, the domain wall has the smallest nonzero growth, and the P4 pattern has the largest initial curvature within the doublon-empty manifold.  At later times, however, the dynamics probes a much larger portion of the Krylov chain.  The growth is then no longer determined only by the first Lanczos coefficient \(b_1\), but by a finite portion of the Lanczos sequence and by the discrete finite-size spectrum of the Krylov Hamiltonian. Consequently, the active-link count does not carry the spread physics.  In a finite system, \(K(t)\) does not relax to a strictly time-independent value.  Instead, after the initial growth regime it oscillates around a state- and parameter-dependent mean.  It is therefore useful to characterise the late-time dynamics not by a putative saturation value, but by a time-window average and a fluctuation scale.

We define the late-time averaged Krylov complexity as
\begin{equation}
\overline K_{\rm late}(\epsilon,\Delta)
=
\frac{1}{T_2(\Delta)-T_1(\Delta)}
\int_{T_1(\Delta)}^{T_2(\Delta)}
K(t;\epsilon,\Delta)\,dt .
\end{equation}
The corresponding fluctuation scale is measured by
\begin{eqnarray}
\sigma_{K,{\rm late}}(\epsilon,\Delta)
=
\Big[
\frac{1}{T_2(\Delta)-T_1(\Delta)}
&&\int_{T_1(\Delta)}^{T_2(\Delta)}
\big(
K(t;\epsilon,\Delta) \nonumber\\
&& -\overline K_{\rm late}(\epsilon,\Delta)
\big)^2 dt
\Big]^{1/2}.
\end{eqnarray}
The same time window is used for both quantities.

The choice of the averaging window needs some care.  It must be late compared to the initial quadratic regime for all non-frozen states included in the comparison.  We choose it using the slowest non-frozen state among the representative states, namely the doublon domain wall.  Here $L$ denotes the total number of rungs, each carrying the two fermionic modes $\alpha_j$ and $\beta_j$; each rung-parity sector contains $L/2$ rungs. Recall, the domain wall has the minimal nonzero active-link count, $A_{\rm DW}=4,$ independent of \(L\) for periodic boundary conditions.  Therefore its first Krylov coefficient obeys
\begin{equation}
b_{1,{\rm DW}}^2
=
16\Delta^2 A_{\rm DW}
=
64\Delta^2,
\qquad
b_{1,{\rm DW}}=8|\Delta|.
\end{equation}
We define the late-time window in units of this slowest growth scale:
\begin{equation}
T_1(\Delta)=\frac{s_1}{8|\Delta|},
\qquad
T_2(\Delta)=\frac{s_2}{8|\Delta|}.
\end{equation}
In the numerical data shown below we use
\begin{equation}
s_1=8,\qquad s_2=16,
\end{equation}
so that
\begin{equation}
T_1(\Delta)=\frac{1}{|\Delta|},
\qquad
T_2(\Delta)=\frac{2}{|\Delta|}.
\end{equation}
Equivalently, the domain-wall state is averaged over the interval
\begin{equation}
b_{1,{\rm DW}}t\in[8,16],
\end{equation}
which is well beyond the initial quadratic regime.  Since the all-\(\beta\)-filled and P4 states have larger first Krylov coefficients, the same window places them even deeper into their finite-size late-time oscillatory regime.  The point \(\Delta=0\) is treated separately: in that limit all the initial states considered here are eigenstates of \(H_{\rm C}\), and the Krylov complexity is time independent.
\begin{figure*}[t]
\centering
\begin{minipage}{0.98\textwidth}
\centering
\textbf{(a)}\par\vspace{2pt}
\includegraphics[width=1.0\textwidth]{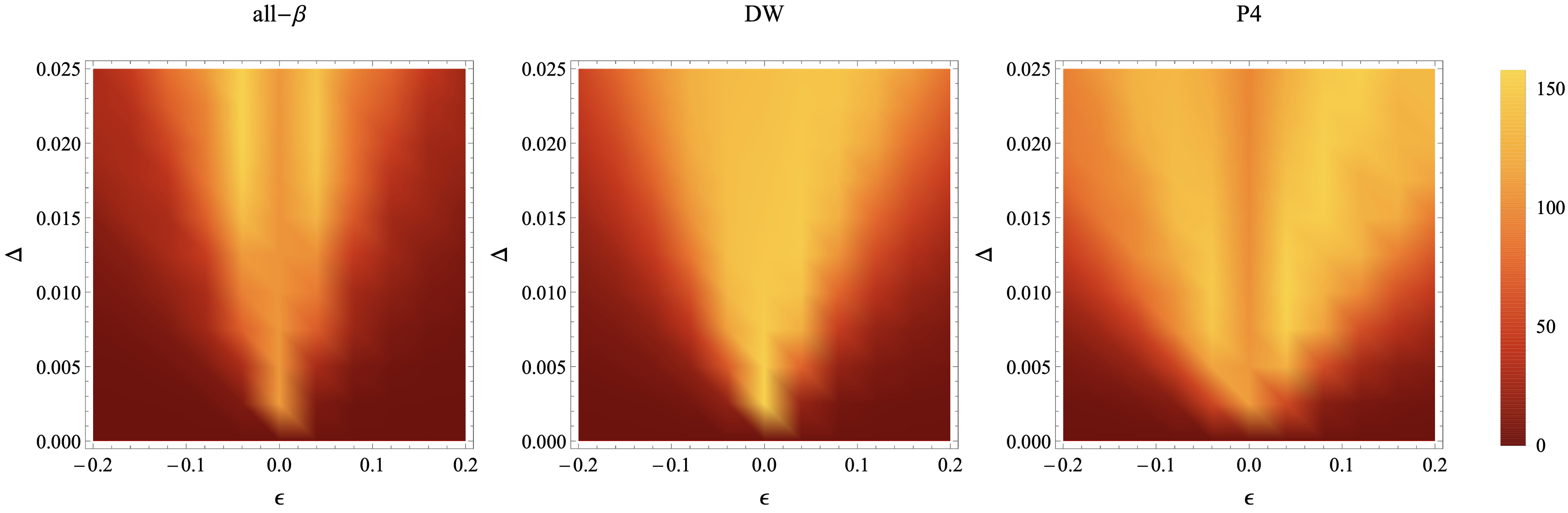}
\end{minipage}
\vspace{0.8em}
\begin{minipage}{0.98\textwidth}
\centering
\textbf{(b)}\par\vspace{2pt}
\includegraphics[width=1.0\textwidth]{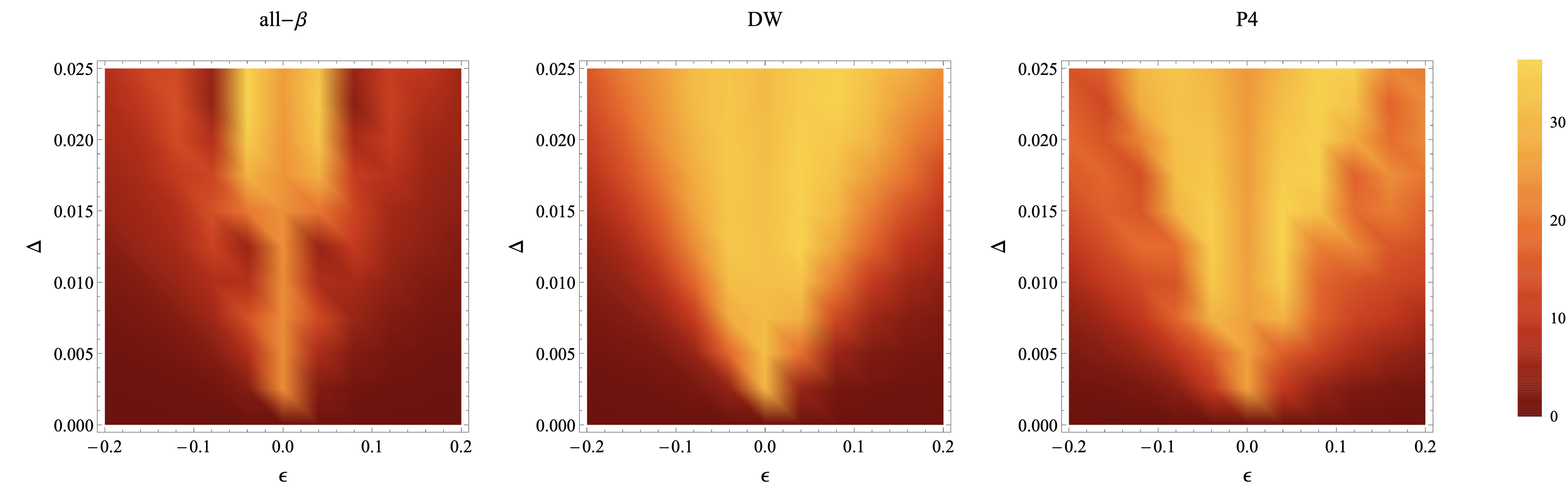}
\end{minipage}
\caption{Late-time finite-size Krylov response in the $(\Delta,\epsilon)$ plane for $L=12$ rungs, ie. 24 sites. Panel (a) shows the window-averaged late-time Krylov complexity $\overline K_{\rm late}$, while panel (b) shows the corresponding late-time fluctuation scale $\sigma_{K,\rm late}$ over the same averaging window.  In each panel the three columns correspond, from left to right, to the all-$\beta$-filled state, the doublon domain-wall state, and the P4 doublon pattern. The panel (b) quantifies the standard deviation of these residual fluctuations about the average values. The narrow dark notch on the exact resonant line \(\epsilon=V+4\tau=0\), most prominent in the all-\(\beta\)
and P4 panels, is a finite-size coherent-resonance feature; see the discussion in the text.}
\label{fig:late_heatmaps}
\end{figure*}

Figure~\eqref{fig:late_heatmaps} shows the resulting late-time response maps.  The upper panel gives \(\overline K_{\rm late}\), while the lower panel shows \(\sigma_{K,{\rm late}}\) computed over the same window.  These plots should be interpreted as finite-size late-time probes, not thermodynamic saturation values.  The mean maps quantify how far, on average, each initial state spreads in Krylov space after the early-growth regime, while the fluctuation maps measure the residual oscillations around that mean.  The CDW state is omitted because it is exactly annihilated by \(H_\Delta\) and remains frozen for all \((\epsilon,\Delta)\). 

A small but useful feature of Figure~\eqref{fig:late_heatmaps} is the narrow darker notch pinned to the
exact resonant line \(\epsilon=V+4\tau=0\), most clearly visible for the all-\(\beta\) and P4
initial states.  Exactly on this line, the off-site particle-hole states reached by \(H_\Delta\) become
degenerate under \(H_{\rm C}\).  The resulting dynamics is therefore coherent. The initial
state couples strongly to a small number of `bright' linear combinations inside the resonant
manifold and exhibits regular oscillations and partial revivals \footnote{On resonant line every same-parity pair $(j\!-\!1,j\!+\!1)$ of the doublon--empty states
$|D0\rangle,|0D\rangle$ are degenerate with the $\beta$--singlon pair $|\beta\beta\rangle$.  The $H_\Delta$ therefore reduces, in the resonant sector, to coherent local
conversions $D0/0D\leftrightarrow \beta\beta$ with a bright/dark structure: one linear combination
(the ``bright'' mode) undergoes oscillations while the orthogonal (``dark'') combination
decouples.}. This is seemingly absent in the DW panel since it only has 4 non-zero matrix elements localised near the two interfaces and hence much lower chance of revivals, thereby coupling more efficiently to larger connected components of the Hilbert space. However, even so slightly away from the line,
the same channels remain near-resonant, but the exact degeneracy is lifted.  The corresponding
Lanczos spectrum becomes more incommensurate, producing stronger dephasing over the averaging
window and hence a larger value of \(\overline K_{\rm late}\).  The dark line is therefore best
understood as a finite-size anti-resonance inside a broader near-resonant response band, rather
than as a true suppression of Carroll-breaking dynamics.

\section{Comparing with a field theory example}\label{secV}

The lattice results above show that Carroll-breaking perturbations immediately generate
Hilbert-space spreading depending on characteristics of the initial state. The corresponding early-time response is controlled by
the state dependent geometric invariants associated to the perturbation acting on otherwise ultra-local states, while the late time response is much more involved.  In this section we
present a continuum analogue of this mechanism for a simple Carrollian scalar field theory.  

The purpose of this calculation is twofold:
(i) to illustrate in a controlled continuum setting the inevitability of Krylov growth
upon breaking Carroll symmetry, and (ii) to clarify the origin of ultraviolet sensitivity
in such growth, which we interpret as a form of UV/IR mixing \cite{cotler2025finite} characteristic of ultra-local
theories.

\subsection{Carroll scalar and gradient deformation}

We consider a free real scalar field theory in one spatial dimension with Hamiltonian
\begin{equation}
H_0=\frac{1}{2}\int dx\,\big(\pi^2+m^2\phi^2\big),
\label{eq:H0_app}
\end{equation}
where the Hamiltonian is understood to be normal ordered so that the vacuum energy is set to zero. 
This theory is ultra-local: field operators at distinct spatial points are completely decoupled, since there is no spatial-gradient term. 
With the canonical commutator
\begin{equation}
[\phi(x),\pi(y)]=i\delta(x-y),
\end{equation}
we use the oscillator normalization
\begin{equation}
\phi(t,x)=\frac{1}{\sqrt{2m}}
\left[
a^\dagger(x)e^{imt}+a(x)e^{-imt}
\right],
\label{eq:phi_app}
\end{equation}
with
\begin{equation}
[a(x),a^\dagger(y)]=\delta(x-y),\qquad 
a(x)|0\rangle=0\qquad \forall x .
\end{equation}
The vacuum $|0\rangle$ is therefore a direct product of local oscillator vacua and contains no spatial entanglement. The equal-time Carroll propagator is strictly local in space,
\begin{equation}
    \langle 0|\phi(x)\phi(y)|0\rangle
    =
    \frac{1}{2m}\delta(x-y).
    \label{eq:carroll-delta-prop}
\end{equation}
This delta-function spatial support is the origin of the cutoff sensitivity below.

The theory in \eqref{eq:H0_app} is the massive version of the electric Carroll scalar theory \cite{deBoer:2021jej,Henneaux:2021yzg}. The mass term does not spoil Carroll symmetry \cite{banerjee2023one}.
\footnote{This theory should not be viewed as the direct continuum limit of the flat-band Creutz ladder. Rather, it provides a continuum example of the same general mechanism: an ultra-local Carrollian system is perturbed by a spatial-gradient deformation.}

To break Carroll symmetry and introduce spatial dynamics, we perturb the system by the relevant deformation
\begin{equation}
H_\lambda=\frac{\lambda}{2}\int dx\,(\partial_x\phi)^2.
\label{eq:Hlambda_app}
\end{equation}
Any perturbation that introduces spatial gradients, or hopping in a lattice realization, immediately couples different points of an otherwise decoupled Carrollian system. We are interested in the Krylov complexity generated by the full Hamiltonian
\begin{equation}
H=H_0+H_\lambda
\end{equation}
when the initial state is the Carroll vacuum $|0\rangle$. This is directly parallel to the lattice problem discussed above, where the Carroll-breaking perturbation is the only source of spreading.

The first Lanczos coefficient is
\begin{equation}
b_1^2
=
\langle 0|
\big(H-\langle H\rangle_0\big)^2
|0\rangle,
\qquad
\langle H\rangle_0\equiv \langle 0|H|0\rangle .
\label{eq:b1def_app}
\end{equation}
Since the normal-ordered $H_0$ annihilates the vacuum, the only nontrivial contribution to the first Krylov step comes from the fluctuation of $H_\lambda$,
\begin{equation}
b_1^2
=
\langle 0|
\big(H_\lambda-\langle H_\lambda\rangle_0\big)^2
|0\rangle .
\label{eq:b1def_lambda_app}
\end{equation}

We now place the system on a circle of circumference $L=2\pi R$ and use the Fourier convention
\begin{equation}
\phi(x)=\frac{1}{\sqrt{L}}\sum_k \phi_k e^{ikx},
\qquad
k=\frac{n}{R},\quad n\in\mathbb{Z}.
\end{equation}
For a real scalar field, $\phi_k^\dagger=\phi_{-k}$, and the oscillator expansion is
\begin{equation}
\phi_k=\frac{1}{\sqrt{2m}}
\left(a_k+a^\dagger_{-k}\right),
\qquad
[a_k,a_q^\dagger]=\delta_{kq}.
\end{equation}
Using
\begin{equation}
\int dx\,(\partial_x\phi)^2
=
\sum_k k^2\phi_k\phi_{-k},
\end{equation}
the perturbation becomes
\begin{equation}
H_\lambda
=
\frac{\lambda}{4m}
\sum_k k^2
\left(
a_k a_{-k}
+a_k a_k^\dagger
+a_{-k}^\dagger a_{-k}
+a_{-k}^\dagger a_k^\dagger
\right).
\label{eq:Hlambda_modes_app}
\end{equation}
The terms proportional to $a_k a_k^\dagger$ produce the vacuum expectation value. After subtracting $\langle H_\lambda\rangle_0$, the only term that acts nontrivially on the vacuum is the pair-creation term:
\begin{equation}
\big(H_\lambda-\langle H_\lambda\rangle_0\big)|0\rangle
=
\frac{\lambda}{4m}
\sum_k k^2
a_{-k}^\dagger a_k^\dagger |0\rangle .
\label{eq:Hlambda_on_vac_app}
\end{equation}
Therefore
\begin{equation}
b_1^2
=
\frac{\lambda^2}{16m^2}
\sum_{k,q} k^2 q^2
\langle 0|
a_k a_{-k}
a^\dagger_{-q}a^\dagger_q
|0\rangle .
\end{equation}
Using
\begin{equation}
\langle 0|
a_k a_{-k}
a^\dagger_{-q}a^\dagger_q
|0\rangle
=
\delta_{kq}+\delta_{k,-q},
\end{equation}
and noting that the $k=0$ mode does not contribute because of the explicit factor $k^2$, we obtain
\begin{equation}
b_1^2
=
\frac{\lambda^2}{8m^2}\sum_k k^4.
\label{eq:b1_sum_app}
\end{equation}
The quartic momentum weight is the direct consequence of the two spatial derivatives in the perturbation.

Introducing a symmetric momentum cutoff $|k|\leq \Lambda$, the mode sum is
\begin{equation}
\sum_{|k|\leq \Lambda} k^4
\simeq
\frac{L}{2\pi}\int_{-\Lambda}^{\Lambda} dk\,k^4
=
\frac{L}{5\pi}\Lambda^5.
\end{equation}
Hence
\begin{equation}
b_1^2
=
\frac{\lambda^2 L}{40\pi m^2}\,\Lambda^5.
\label{eq:b1final_app}
\end{equation}

At early times, the Krylov complexity has the universal expansion
\begin{equation}
C_K(t)=b_1^2t^2+O(t^4),
\end{equation}
so in the present case
\begin{equation}
C_K(t)
=
\frac{\lambda^2 L}{40\pi m^2}\,\Lambda^5\,t^2
+O(t^4).
\label{eq:CK_app}
\end{equation}
Thus the leading growth is quadratic both in time and in the Carroll-breaking coupling $\lambda$. The important point is that the coefficient is dominated by ultraviolet modes.

One can do the same calculation in $d_s$ spatial dimensions, where
\begin{equation}
H_\lambda=\frac{\lambda}{2}\int d^{d_s}x\,(\nabla\phi)^2,
\end{equation}
gives
\begin{equation}
    b_1^2
    =
    \frac{\lambda^2}{8m^2}
    V_{d_s}
    \int_{|\mathbf{k}|\leq \Lambda}
    \frac{d^{d_s}k}{(2\pi)^{d_s}}\,
    |\mathbf{k}|^4
    =
    \frac{\lambda^2 V_{d_s}S_{d_s-1}}
    {8m^2(2\pi)^{d_s}(d_s+4)}
    \Lambda^{d_s+4},
    \label{eq:b1-general-d}
\end{equation}
where $V_{d_s}$ is the spatial volume and $S_{d_s-1}$ is the area of the unit $(d_s-1)$-sphere.

Equivalently, the early-time Krylov growth is controlled by the second moment of the spectral measure associated with the initial state,
\begin{equation}
C_K(t)=\mu_2 t^2+\cdots,
\qquad
\mu_2=
\langle 0|
\big(H-\langle H\rangle_0\big)^2
|0\rangle
=b_1^2 .
\end{equation}
More generally, the moments can be extracted from the autocorrelation function \cite{avdoshkin2024krylov,dymarsky2021krylov}. With
\begin{equation}
C(t)=\langle 0|e^{-i(H-\langle H\rangle_0)t}|0\rangle,
\end{equation}
one may write
\begin{equation}
\mu_{2n}
=
(-1)^n
\frac{d^{2n}}{dt^{2n}}C(t)\Big|_{t=0},
\qquad n\in\mathbb{Z}^+ .
\end{equation}
For the present gradient quench, the expression above reproduces \eqref{eq:b1final_app} for $n=1$.
\subsection{Ultraviolet sensitivity and UV/IR mixing}

The appearance of a strong ultraviolet divergence in \eqref{eq:CK_app} may appear
surprising, as Krylov complexity at early times might be naively regarded as a low-energy
quantity.  This ultraviolet sensitivity is, however, a direct and physical consequence of
the ultra-local nature of the Carrollian vacuum. In the unperturbed theory \eqref{eq:H0_app}, spatial points are completely decoupled and
correlation functions are strictly local with a spatial delta function.  As a result, the vacuum does not suppress
high-momentum fluctuations.  When the gradient deformation \eqref{eq:Hlambda_app} is
introduced, it couples neighboring points and excites arbitrarily short-wavelength modes,
rendering infrared observables sensitive to the ultraviolet cutoff. The power-law sensitivity to the UV cutoff can be traced back to the unperturbed propagator in \eqref{eq:carroll-delta-prop} which has no spatial
falloff: all spatial structure is compressed into the delta function.  Composite operators built
from \(\phi\), and in particular from spatial derivatives of \(\phi\), therefore probe powers of
\(\delta(0)\), or equivalently powers of the momentum cutoff.

In this precise sense, the gradient quench exhibits a Carrollian form of UV/IR mixing.  The
observable \(C_K(t)\) is an infrared dynamical diagnostic of the initially ultra-local theory:
it measures the spreading of a spatially homogeneous vacuum under a weak, relevant deformation.
Nevertheless, its leading coefficient is dominated by arbitrarily short wavelengths,
\begin{equation}
    K(t)\propto \Lambda^5 t^2
    \qquad (d_s=1).
\end{equation}
Thus the Carroll limit obstructs ordinary scale separation.  A low-energy dynamical quantity
cannot be assigned a universal continuum value unless the deformation strength is scaled with
the cutoff.

Indeed, if one insists on a finite continuum limit of the early-time Krylov curvature, in arbitrary spatial dimensions one must
scale
\begin{equation}
    \lambda^2 \Lambda^{d_s+4}
    =
    \text{fixed}.
    \label{eq:lambda-cutoff-scaling}
\end{equation}
This is a double-scaling prescription rather than an ordinary Wilsonian decoupling of UV
physics.  It is analogous in spirit to the cutoff-sensitive structure of Carrollian field
theories, where the number of degrees of freedom per spatial cell enters long-distance
quantities and a finite effective theory requires an additional scaling prescription.

For the lattice flat-band systems studied in the earlier sections, the UV cutoff is physical: it is
set by the inverse lattice spacing.  The divergence in \eqref{eq:CK_app} is therefore
replaced by a finite but lattice-scale-sensitive Krylov curvature.  This is why the continuum
calculation above should be read as a complementary presentation of the same mechanism as that of the Creutz ladder on a lattice.  The common physical point is that ultra-local Carroll
states are exceptionally sensitive to perturbations that restore spatial gradients or hopping.

%This phenomenon constitutes a form of ultraviolet/infrared (UV/IR) mixing characteristic of Carroll-symmetric theories: infrared dynamical quantities, such as early-time Krylov growth, inherit ultraviolet sensitivity due to the absence of spatial correlations in the Carroll limit. It should be emphasised that this behavior does not signal a pathology of the theory or of the Krylov construction.  Rather, it reflects the extreme sensitivity of ultra-local Carrollian systems to spatial-gradient perturbations. The ultraviolet scale $\Lambda$ in the continuum calculation plays a role analogous to the inverse lattice spacing in flat-band lattice models, where compact localised states and zero group velocity similarly lead to enhanced sensitivity to perturbations that restore spatial dynamics.

The analysis presented here is intended to illustrate the universal early-time behavior of
Krylov complexity in Carrollian systems.  It provides a transparent
field-theoretic perspective on the mechanisms underlying the numerical results discussed
in the previous sections. In both given cases, an IR observable – the early‑time growth rate of Krylov complexity is determined entirely by UV/microscopic details. In this sense exact microscopic configuration of doublons as encoded in the link invariant $A$ (which could be $\sim L$), is the discrete analogue of $\sum_k k^4$, both of which counts how many opportunities does the perturbation have to populate all the shortest distance modes available. The comparison, however naive, can be justified since both the objects are a direct consequence of the exact form of the perturbation.

\section{Discussion and Conclusion}\label{secend}
In this work, we offer Krylov spread complexity as a novel probe for breaking of flat-bands structures under perturbations. This becomes possible as we quantify the emergence of flat-bands as direct consequence of Carrollian supertranslations. Subsequently, this means destruction of flat-bands occur due to addition of Carroll relevant deformations to the Hamiltonian. We achieved the same breaking of flat-bands via quenching our supertranslation invariant lattice Hamiltonian in an appropriate way, choosing initial states in particular phases of the parameter space. We also supplemented this with a continuum Carroll scalar field calculation, where the Carroll breaking is achieved by a gradient perturbation.

Our lattice model results establish Krylov complexity as a sensitive dynamical probe of Carroll
symmetry protection and provide a many-body, basis-independent diagnostic of the
fragility or robustness of Carroll-symmetric phases. A special regime of interest is the exotic phase, which is hugely degenerate, and we quantified the distinct opportunities of different states to spread under the same perturbation using active-link invariants. Such a discussion of how the formerly equal ground states fracture into distinct dynamical classes pertaining to Krylov spreading on an initial state manifold makes the flat-band breaking very special. In other words, Krylov in this case is an excellent probe as it is sensitive to the spectral measure of the seed.

For the continuum electric Carroll scalar, the analogous ingredient is the flatness of the excitation spectrum and the ultra-local structure of the vacuum. The absence of spatial gradients makes the vacuum correlation function proportional to a spatial delta function, so a gradient deformation probes arbitrarily short
wavelengths immediately. The early time growth in this case, intriguingly, varies directly with the UV momentum cutoff.

This unique structure of spread complexity for Carroll theories leave open a multitude of directions to investigate upon in the future. Since at the end of the day we are interested in the dynamics of Lanczos coefficients, the universality classes of $b_n$ for generic Carroll models across geometry and dimension is of utmost importance to us. One could try to understand the nuances associated to Krylov dynamics in such systems more closely, perhaps with different perturbations, that may enhance/suppress growth of certain states w.r.t others. 

An even more pertinent question is to look at Krylov dynamics is low energy effective continuum models, due to this UV/IR mixing alluded to here. Flat-bands augmented by supertranslation invariance naturally appears in systems like twisted bilayer graphene at magic angles \cite{Bagchi:2022eui}, for which the effective field theory is well understood \cite{tarnopolsky2019origin}. It might be very interesting to understand Krylov spread in this model tuned away from the magic angles. Further, there are other classes of Carroll invariant scalar field theories, namely the \textit{magnetic} ones involving spatial derivatives \cite{Henneaux:2021yzg}, for which the logic of Krylov spread above needs to be adjusted from first principles!

With a multitude of pressing questions in hand, we thus hope to return to this line of investigations in future installments.

\section*{Acknowledgements:}
The authors would like to thank the participants in SSQDT in BITS Pilani KK Birla Goa Campus, where parts of this work were presented and discussed. ABan is supported in part by an OPERA grant and a seed grant NFSG/PIL/2023/P3816 from BITS-Pilani, and further an early career research grant ANRF/ECRG/2024/002604/PMS from ANRF India. He also acknowledges financial support from the Asia Pacific Center for Theoretical Physics (APCTP) via an Associate Fellowship. He thanks Department of Theoretical Physics, Saitama University, for kind hospitality when this work was finalised.  AB is supported by the Core Research Grant (CRG/2023/ 001120) by the Department of Science and Technology, Science and Engineering Research Board (India), India. AB also acknowledges the associateship program of the Indian Academy of Science, Bengaluru, and the generous donor through the Singheswari and Ram Krishna Jha Chair. Research of RB is supported partially by an intramural grant CDRF C2/24/282 from BITS Pilani and ANRF India through ANRF/ARG/2025/009241/PS. He thanks the generous hospitality of APCTP, Pohang, during the workshop "Field Theory and Geometry in Fractional Quantum Hall and Flat Band Systems", where this work was partially carried out and discussed.

\appendix

\section{Brief introduction to Carroll symmetry}
The $(d+1)$-dimensional Carroll algebra is generated by spatial
rotations $J_{ij}$, spatial translations $P_i$, Hamiltonian $H$, and Carroll boost generators $C_i$.
Starting from the Poincar\'e algebra, the Lorentz boost generators, when all $c$ factors are reinstated, are
given by \cite{deBoer:2021jej}
\begin{equation}
L_i=\frac{1}{c}x^i\partial_t+ct\,\partial_i .
\end{equation}
The Carrollian contraction is then obtained through scaling and contracting the boost generators and the time translation $P_0$:
\begin{equation}
C_i=cL_i,
\qquad
H=cP_0,
\qquad
c\rightarrow0 .
\end{equation}
The rotation generators remain unchanged in this contraction.
Substituting the expression for $L_i$, one finds
\begin{equation}
C_i=x^i\partial_t+c^2 t\,\partial_i .
\end{equation}
In the ultra-relativistic limit $c\to0$, this reduces to
\begin{equation}
C_i\rightarrow x^i\partial_t ,
\end{equation}
which defines the Carroll boost generator.
The resulting non-vanishing commutation relations are
\begin{align}
[J_{ij},J_{kl}] &=
\delta_{jk}J_{il}
-\delta_{ik}J_{jl}
-\delta_{jl}J_{ik}
+\delta_{il}J_{jk},
\\
[J_{ij},P_k] &= \delta_{jk}P_i-\delta_{ik}P_j,
\\
[J_{ij},C_k] &= \delta_{jk}C_i-\delta_{ik}C_j,
\\
[C_i,P_j] &= \delta_{ij}H .
\end{align}
Interestingly, Carroll boosts commute: $[C_i, C_j]=0$, making this a crucial difference from the Poincaré algebra. Interestingly, one can have an infinite extension of the boost generators, $M_i = f(x_i)\partial_t$ with some smooth function $f$ so that changing $C_i \to M_i$ does not change the resultant algebra. These vector fields $M_i$ constitute supertranslations. 

\section{Counting degenerate states}
\label{statecount}
The Hamiltonian for our interacting supertranslation invariant system is given by:
\begin{equation} \label{orig}
    H_0 = \sum_{j=1}^L \left(V n^{\alpha}_j n^{\beta}_j + 2\tau (n^{\alpha}_j - n^{\beta}_j) \right)
\end{equation}
Since this Hamiltonian is purely on-site (diagonal in the rung occupation basis), its eigenstates are product states of the $L$ rungs. We can find the full spectrum by analyzing the four possible states of a single rung and applying the system-wide constraints.

The energy for a single rung $j$ depends on its four possible occupation states:
\begin{itemize}
    \item $E(0,0)$ for $n^\alpha=0, n^\beta=0$ (empty): $E_{00} = 0$
    \item $E(1,0)$ for $n^\alpha=1, n^\beta=0$ ($\alpha$-filled): $E_{10} = 2\tau$
    \item $E(0,1)$ for $n^\alpha=0, n^\beta=1$ ($\beta$-filled): $E_{01} = -2\tau$
    \item $E(1,1)$ for $n^\alpha=1, n^\beta=1$ (doublon): $E_{11} = V$
\end{itemize}

Any eigenstate of the full $L$-rung system can be characterized by the number of rungs in each of these four configurations. Let us define $\mathfrak{g}_{00}$, $\mathfrak{g}_{10}$, $\mathfrak{g}_{01}$, and $\mathfrak{g}_{11}$ as the integer counts of rungs in each of these states. In the half-filled sector, it imposes two strict constraints on these counts:
\begin{enumerate}
    \item Total Rung Constraint: The total number of rungs must be $L$.
    \begin{equation} \label{constr1}
        \mathfrak{g}_{00} + \mathfrak{g}_{10} + \mathfrak{g}_{01} + \mathfrak{g}_{11} = L
    \end{equation}
    \item Half-Filling Constraint: The total number of particles (on $2L$ sites) must be $L$.
    \begin{equation} \label{constr2}
        0\cdot \mathfrak{g}_{00} + 1\cdot \mathfrak{g}_{10} + 1\cdot \mathfrak{g}_{01} + 2\cdot \mathfrak{g}_{11} = L
    \end{equation}
\end{enumerate}
These two constraints lead to the determination of the entire spectrum. By subtracting Equation \eqref{constr1} from Equation \eqref{constr2}, we find:
\begin{equation}
    \mathfrak{g}_{11} = \mathfrak{g}_{00}
\end{equation}
This shows that any allowed state in the half-filled sector must have an equal number of doubly-occupied rungs and empty rungs.

The above constraint allows us to organise the entire eigenspectrum into ``bands" indexed by a single integer $\mathfrak{g} \equiv \mathfrak{g}_{11} = \mathfrak{g}_{00}$. The allowed values for $\mathfrak{g}$ are $\mathfrak{g} \in \{0, 1, \dots, \lfloor L/2 \rfloor\}$, where $\lfloor.\rfloor$ denotes the floor function. Within each $\mathfrak{g}$-band, the remaining $L-2\mathfrak{g}$ rungs are filled by $\mathfrak{g}_{10}$ and $\mathfrak{g}_{01}$ rungs. These bands are further split into distinct energy levels indexed by a second integer $m \equiv \mathfrak{g}_{10}$. The number of $\mathfrak{g}_{01}$ rungs is then fixed as $\mathfrak{g}_{01} = L - 2\mathfrak{g} - m$. The allowed values for $m$ are $m \in \{0, 1, \dots, L-2\mathfrak{g}\}$. Thus, any distinct energy level in the system can be uniquely identified by the pair of integers $(\mathfrak{g}, m)$.

We can now write the general formula for the energy $E(\mathfrak{g}, m)$ of any level:
\begin{equation}
    E(\mathfrak{g}, m) = \mathfrak{g} \cdot E_{11} + \mathfrak{g} \cdot E_{00} + m \cdot E_{10} + (L-2\mathfrak{g}-m) \cdot E_{01}
\end{equation}
Substituting the energies:
%\begin{equation}
%    E(\mathfrak{g}, m) = \mathfrak{g}(V) + \mathfrak{g}(0) + m(2\tau) + (L-2\mathfrak{g}-m)(-2\tau)
%\end{equation}
\begin{equation} \label{energy}
    E(\mathfrak{g}, m) = %\mathfrak{g}V + 2m\tau - 2L\tau + 4\mathfrak{g}\tau + 2m\tau = 
    \mathfrak{g}(V + 4\tau) + m(4\tau) - 2L\tau
\end{equation}
%The total energy for any level $(\mathfrak{g},m)$ in the spectrum is:
%\begin{equation}
%    E(\mathfrak{g}, m) = \mathfrak{g}(V + 4\tau) + m(4\tau) - 2L\tau
%\end{equation}
The degeneracy $D(\mathfrak{g}, m)$ of this level is the number of ways to arrange these rungs on the $L$-rung lattice. This is a classic combinatorial problem:
\begin{equation}
    D(\mathfrak{g}, m) = \binom{L}{\mathfrak{g}} \binom{L-\mathfrak{g}}{\mathfrak{g}} \binom{L-2\mathfrak{g}}{m}
\end{equation}
This can be written more compactly as the multinomial coefficient:
\begin{equation}
    D(\mathfrak{g}, m) = \frac{L!}{\mathfrak{g}!  \mathfrak{g}!  m!  (L-2\mathfrak{g}-m)!}
\end{equation}
%This formula gives the exact degeneracy for every energy level $E(\mathfrak{g}, m)$.

A special physical regime occurs at the critical point $V = -4\tau$. If we substitute this value into our energy formula \eqref{energy}, the entire $\mathfrak{g}$-dependence of the spectrum vanishes:
%\begin{equation}
%    E(\mathfrak{g}, m) \text{ at } V=-4\tau = \mathfrak{g}(-4\tau + 4\tau) + m(4\tau) - 2L\tau
%\end{equation}
\begin{equation}
    E(m) = 4\tau m - 2L\tau \quad (\text{at } V=-4\tau)
\end{equation}
At this critical point, the energy only depends on the quantum number $m$ (the number of $\alpha$-filled rungs). All states with different $\mathfrak{g}$-values but the same $m$-value merge into a single, massively degenerate energy level. This creates a shattered spectrum of $L+1$ distinct, equally-spaced energy levels, $E(m)$, separated from each other by a large, macroscopic gap of $4\tau$. The total degeneracy of each of these merged levels $E(m)$ is the sum of all their constituent degeneracies:
\begin{equation}
    D_m = \sum_{\mathfrak{g}=0}^{\lfloor (L-m)/2 \rfloor} D(\mathfrak{g}, m) = \sum_{\mathfrak{g}=0}^{\lfloor (L-m)/2 \rfloor} \frac{L!}{\mathfrak{g}! \cdot \mathfrak{g}! \cdot m! \cdot (L-2\mathfrak{g}-m)!}
\end{equation}
%This structure is the key to understanding the localization and Krylov shattering observed in numerical simulations. 
For example, the low-energy subspace $m=0$, which is gapped from the $m=1$ subspace by $4\tau$, has a total degeneracy of:
\begin{equation}
    D_0 = \sum_{\mathfrak{g}=0}^{\lfloor L/2 \rfloor} \frac{L!}{\mathfrak{g}! \cdot \mathfrak{g}! \cdot (L-2\mathfrak{g})!}
\end{equation}
This subspace, which scales as $\sim 3^L$ (as opposed to the full $\sim 4^L$ Hilbert space), acts as an isolated low-energy laboratory for perturbations at or near $V = -4\tau$.

\section{More on active link dynamics} \label{Prop_App}
A major result of this paper is counting of active links associated to the initial states in the degenerate vacuum manifold for the Exotic phase. In what follows, we will describe some salient features of this idea using different inital configurations.

%\begin{widetext}
%    \begin{minipage}{\linewidth}
           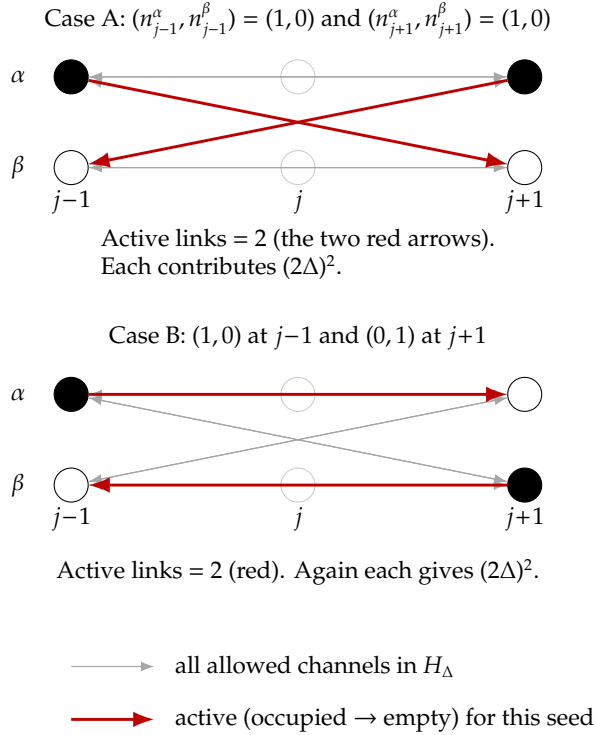
\begin{figure}[htb]
\centering
\begin{tikzpicture}[
  >=Latex,
  site/.style={circle,draw,minimum size=4.5mm,inner sep=0pt},
  occ/.style ={site,fill=black},
  emp/.style ={site,fill=white},
  allhop/.style={->,draw=black!35,line width=0.45pt},
  active/.style={->,draw=red!75!black,line width=1.1pt},
  lab/.style={font=\small}
]

% ---------- helper: draw a 3-rung "cell" with labels ----------
\newcommand{\cell}[4]{%
  % #1 = xshift, #2 = title, #3/#4 = occupancy pattern selector
  \begin{scope}[xshift=#1]
    \node[lab] at (3,1.35) {#2};

    % rung positions
    \node[lab] at (0, -1.05) {$j\!-\!1$};
    \node[lab] at (3, -1.05) {$j$};
    \node[lab] at (6, -1.05) {$j\!+\!1$};

    % alpha/beta labels
    \node[lab] at (-0.7,  0.6) {$\alpha$};
    \node[lab] at (-0.7, -0.6) {$\beta$};

    % sites (we only use j-1 and j+1 for H_\Delta hops; j is just a label)
    % left rung (j-1)
    \node[#3] (aL) at (0,  0.6) {};
    \node[#4] (bL) at (0, -0.6) {};

    % middle rung (j) (shown faintly)
    \node[site,draw=black!20] (aM) at (3,  0.6) {};
    \node[site,draw=black!20] (bM) at (3, -0.6) {};

    % right rung (j+1)
    \node (aR) at (6,  0.6) {};
    \node (bR) at (6, -0.6) {};
  \end{scope}
}

% --- Panel A: (j-1) and (j+1) same flavor occupied: alpha-alpha around j ---
% Occupancies:
%  j-1: alpha occupied, beta empty
%  j+1: alpha occupied, beta empty
\begin{scope}
  \node[lab] at (0,2.15) {};
\end{scope}

\begin{scope}[xshift=0cm]
  \node[lab] at (3,1.35) {Case A: $(n_{j-1}^\alpha,n_{j-1}^\beta)=(1,0)$ and $(n_{j+1}^\alpha,n_{j+1}^\beta)=(1,0)$};

  % sites
  \node[occ] (aL1) at (0,  0.6) {};
  \node[emp] (bL1) at (0, -0.6) {};
  \node[site,draw=black!20] (aM1) at (3,  0.6) {};
  \node[site,draw=black!20] (bM1) at (3, -0.6) {};
  \node[occ] (aR1) at (6,  0.6) {};
  \node[emp] (bR1) at (6, -0.6) {};

  \node[lab] at (0, -1.05) {$j\!-\!1$};
  \node[lab] at (3, -1.05) {$j$};
  \node[lab] at (6, -1.05) {$j\!+\!1$};
  \node[lab] at (-0.7,  0.6) {$\alpha$};
  \node[lab] at (-0.7, -0.6) {$\beta$};

  % all possible H_\Delta channels between j-1 and j+1 (thin gray, both directions)
  \draw[allhop] (aL1) -- (aR1); \draw[allhop] (aR1) -- (aL1); % alpha-alpha
  \draw[allhop] (bL1) -- (bR1); \draw[allhop] (bR1) -- (bL1); % beta-beta
  \draw[allhop] (aL1) -- (bR1); \draw[allhop] (bR1) -- (aL1); % alpha->beta
  \draw[allhop] (bL1) -- (aR1); \draw[allhop] (aR1) -- (bL1); % beta->alpha

  % Active oriented moves (occupied source -> empty target):
  % Here: aL occupied, bR empty  => aL -> bR active  (term beta_{j+1}^\dagger alpha_{j-1})
  %       aR occupied, bL empty  => aR -> bL active  (term beta_{j-1}^\dagger alpha_{j+1} = h.c. of alpha_{j+1}^\dagger beta_{j-1})
  \draw[active] (aL1) -- (bR1);
  \draw[active] (aR1) -- (bL1);

  \node[lab,align=left] at (3,-1.75)
  {Active links = 2 (the two red arrows).\\ Each contributes $(2\Delta)^2$.};
\end{scope}

% --- Panel B: opposite flavors occupied around j (domain-wall neighborhood) ---
\begin{scope}[xshift=0cm,yshift=-4.2cm]
  \node[lab] at (3,1.35) {Case B: $(1,0)$ at $j\!-\!1$ and $(0,1)$ at $j\!+\!1$};

  % sites
  \node[occ] (aL2) at (0,  0.6) {};
  \node[emp] (bL2) at (0, -0.6) {};
  \node[site,draw=black!20] (aM2) at (3,  0.6) {};
  \node[site,draw=black!20] (bM2) at (3, -0.6) {};
  \node[emp] (aR2) at (6,  0.6) {};
  \node[occ] (bR2) at (6, -0.6) {};

  \node[lab] at (0, -1.05) {$j\!-\!1$};
  \node[lab] at (3, -1.05) {$j$};
  \node[lab] at (6, -1.05) {$j\!+\!1$};
  \node[lab] at (-0.7,  0.6) {$\alpha$};
  \node[lab] at (-0.7, -0.6) {$\beta$};

  % all possible channels (thin gray, both directions)
  \draw[allhop] (aL2) -- (aR2); \draw[allhop] (aR2) -- (aL2);
  \draw[allhop] (bL2) -- (bR2); \draw[allhop] (bR2) -- (bL2);
  \draw[allhop] (aL2) -- (bR2); \draw[allhop] (bR2) -- (aL2);
  \draw[allhop] (bL2) -- (aR2); \draw[allhop] (aR2) -- (bL2);

  % Active oriented moves (occupied source -> empty target):
  % aL occupied, aR empty => aL -> aR active (term alpha_{j+1}^\dagger alpha_{j-1})
  % bR occupied, bL empty => bR -> bL active (term beta_{j-1}^\dagger beta_{j+1} = h.c. of beta_{j+1}^\dagger beta_{j-1})
  \draw[active] (aL2) -- (aR2);
  \draw[active] (bR2) -- (bL2);

  \node[lab,align=left] at (3,-1.75)
  {Active links = 2 (red). Again each gives $(2\Delta)^2$.};
\end{scope}

% Legend
\begin{scope}[xshift=0.0cm,yshift=-7.2cm]
  \draw[allhop] (0,0) -- (1.1,0);
  \node[lab,anchor=west] at (1.25,0) {all allowed channels in $H_\Delta$};
  \draw[active] (0,-0.7) -- (1.1,-0.7);
  \node[lab,anchor=west] at (1.25,-0.7) {active (occupied $\to$ empty) for this seed};
\end{scope}
\end{tikzpicture}
\caption{Heuristic counting rule visualisation for one fermions per rung: for each center $j$, look at the four channels
connecting $(j-1)\leftrightarrow (j+1)$ in $H_\Delta$; an oriented arrow is ``active'' iff its source
site is occupied and its target site is empty in the chosen Fock seed. Each active oriented link
contributes one term $(2\Delta)^2$ to
$b_1^2$.}
\label{oneferm}
\end{figure}
%    \end{minipage}
%\end{widetext}

We consider the case of our \(H_\Delta\) in \eqref{eq:Hdelta} now.
Let the occupancies in \(|\psi_0\rangle\) be \(n_j^\alpha, n_j^\beta \in \{0,1\}\).
\(H_\Delta\) contains, for each rung \(j\), hopping terms between \(j-1\) and \(j+1\) in four channels:
\begin{equation}
\begin{aligned}
\alpha\to\alpha &: \ \alpha_{j+1}^\dagger \alpha_{j-1} + \text{h.c.},\\
\beta\to\beta &: \ \beta_{j+1}^\dagger \beta_{j-1} + \text{h.c.},\\
\alpha\to\beta &: \ \beta_{j+1}^\dagger \alpha_{j-1} + \text{h.c.},\\
\beta\to\alpha &: \ \alpha_{j+1}^\dagger \beta_{j-1} + \text{h.c.}
\end{aligned}
\end{equation}
A term contributes iff the source orbital is occupied and the target is empty.
Thus the number of active oriented moves across the bond centered at \(j\) for each channel is
\begin{align}
A_j^{\alpha\alpha} &= n_{j-1}^\alpha(1-n_{j+1}^\alpha) + n_{j+1}^\alpha(1-n_{j-1}^\alpha),\nonumber\\
A_j^{\beta\beta}  &= n_{j-1}^\beta(1-n_{j+1}^\beta) + n_{j+1}^\beta(1-n_{j-1}^\beta),\nonumber\\
A_j^{\alpha\beta} &= n_{j-1}^\alpha(1-n_{j+1}^\beta) + n_{j+1}^\beta(1-n_{j-1}^\alpha),\nonumber\\
A_j^{\beta\alpha} &= n_{j-1}^\beta(1-n_{j+1}^\alpha) + n_{j+1}^\alpha(1-n_{j-1}^\beta).
\end{align}
The total number of active moves (i.e. the norm squared of \(H_\Delta|\psi_0\rangle\) up to a factor \(\Delta^2\)) is
\begin{equation}
\mathcal{N}_{\text{active}}(\psi_0) = \sum_{j=1}^L \left( A_j^{\alpha\alpha} + A_j^{\beta\beta} + A_j^{\alpha\beta} + A_j^{\beta\alpha} \right).
\end{equation}
For a toy example, take the case in Figure \eqref{oneferm} with one fermion per rung. Upper panel shows a situation where both neighboring rungs carry an $\alpha$ fermion; then only the cross-channel hops are active, giving two active oriented moves. Panel below shows the domain-wall neighborhood with only $\alpha$ at $j$ and only $\beta$ at $j+1$, then only the same-flavor channels are active, again giving two active oriented moves.

Coming back to our situation of interest, since \(n_j^\alpha = n_j^\beta = d_j\) inside our doublon/empty manifold, the counting here is different, as every channel is non-zero, giving the identical expression
\begin{equation}
A_j^{\mu\nu} = d_{j-1}(1-d_{j+1}) + d_{j+1}(1-d_{j-1}) .
\end{equation}
Using the elementary identity \(d(1-d') + d'(1-d) = (d-d')^2\) for binary variables, we obtain
\begin{equation}
A_j^{\mu\nu} = (d_{j-1} - d_{j+1})^2 = \chi_j .
\end{equation}

Thus the total number of active hopping processes (summed over all bonds and all four channels) is
\begin{equation}
\mathcal{N}_{\text{active}}(\psi_0) = \sum_{j=1}^{L} 4\,(d_{j-1} - d_{j+1})^2 .
\end{equation}
As described in the main text, it completely determines the leading Lanczos coefficient and hence the initial growth of Krylov complexity.

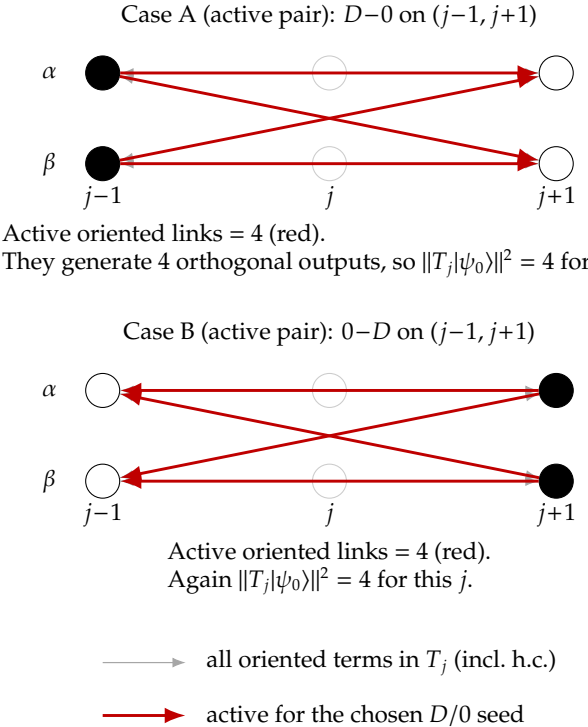
\begin{figure}[htb]
\centering
\begin{tikzpicture}[
  >=Latex,
  site/.style={circle,draw,minimum size=4.5mm,inner sep=0pt},
  occ/.style ={site,fill=black},
  emp/.style ={site,fill=white},
  allhop/.style={->,draw=black!35,line width=0.45pt},
  active/.style={->,draw=red!75!black,line width=1.1pt},
  lab/.style={font=\small}
]

% ---------------- Panel A: D-0 ----------------
\begin{scope}[xshift=0cm]
  \node[lab] at (3,1.35) {Case A (active pair): $D\!-\!0$ on $(j\!-\!1,j\!+\!1)$};

  % sites: rung j-1 (left)
  \node[occ] (aL1) at (0,  0.6) {};
  \node[occ] (bL1) at (0, -0.6) {};
  % middle rung j (faint, just for reference)
  \node[site,draw=black!20] (aM1) at (3,  0.6) {};
  \node[site,draw=black!20] (bM1) at (3, -0.6) {};
  % sites: rung j+1 (right)
  \node[emp] (aR1) at (6,  0.6) {};
  \node[emp] (bR1) at (6, -0.6) {};

  % labels
  \node[lab] at (0, -1.05) {$j\!-\!1$};
  \node[lab] at (3, -1.05) {$j$};
  \node[lab] at (6, -1.05) {$j\!+\!1$};
  \node[lab] at (-0.7,  0.6) {$\alpha$};
  \node[lab] at (-0.7, -0.6) {$\beta$};

  % all oriented channels in T_j between j-1 and j+1 (thin gray, both directions)
  \draw[allhop] (aL1) -- (aR1); \draw[allhop] (aR1) -- (aL1); % alpha-alpha
  \draw[allhop] (bL1) -- (bR1); \draw[allhop] (bR1) -- (bL1); % beta-beta
  \draw[allhop] (aL1) -- (bR1); \draw[allhop] (bR1) -- (aL1); % alpha-beta
  \draw[allhop] (bL1) -- (aR1); \draw[allhop] (aR1) -- (bL1); % beta-alpha

  % Active oriented moves for D-0: four hops from left (occupied) to right (empty)
  %  alpha->alpha, beta->beta, alpha->beta, beta->alpha
  \draw[active] (aL1) -- (aR1); % alpha_{j+1}^\dagger alpha_{j-1}
  \draw[active] (bL1) -- (bR1); % beta_{j+1}^\dagger beta_{j-1} (sign irrelevant for activity)
  \draw[active] (aL1) -- (bR1); % beta_{j+1}^\dagger alpha_{j-1}
  \draw[active] (bL1) -- (aR1); % alpha_{j+1}^\dagger beta_{j-1}

  \node[lab,align=left] at (3,-1.75)
  {Active oriented links = 4 (red).\\
   They generate 4 orthogonal outputs, so $\|T_j|\psi_0\rangle\|^2=4$ for this $j$.};
\end{scope}

% ---------------- Panel B: 0-D ----------------
\begin{scope}[xshift=0cm,yshift=-4.2cm]
  \node[lab] at (3,1.35) {Case B (active pair): $0\!-\!D$ on $(j\!-\!1,j\!+\!1)$};

  % left rung j-1: empty
  \node[emp] (aL2) at (0,  0.6) {};
  \node[emp] (bL2) at (0, -0.6) {};
  % middle rung j (faint)
  \node[site,draw=black!20] (aM2) at (3,  0.6) {};
  \node[site,draw=black!20] (bM2) at (3, -0.6) {};
  % right rung j+1: doublon
  \node[occ] (aR2) at (6,  0.6) {};
  \node[occ] (bR2) at (6, -0.6) {};

  % labels
  \node[lab] at (0, -1.05) {$j\!-\!1$};
  \node[lab] at (3, -1.05) {$j$};
  \node[lab] at (6, -1.05) {$j\!+\!1$};
  \node[lab] at (-0.7,  0.6) {$\alpha$};
  \node[lab] at (-0.7, -0.6) {$\beta$};

  % all oriented channels (thin gray)
  \draw[allhop] (aL2) -- (aR2); \draw[allhop] (aR2) -- (aL2);
  \draw[allhop] (bL2) -- (bR2); \draw[allhop] (bR2) -- (bL2);
  \draw[allhop] (aL2) -- (bR2); \draw[allhop] (bR2) -- (aL2);
  \draw[allhop] (bL2) -- (aR2); \draw[allhop] (aR2) -- (bL2);

  % Active oriented moves for 0-D: now only the h.c. terms act,
  % i.e. four hops from right (occupied) to left (empty)
  \draw[active] (aR2) -- (aL2);
  \draw[active] (bR2) -- (bL2);
  \draw[active] (aR2) -- (bL2);
  \draw[active] (bR2) -- (aL2);

  \node[lab,align=left] at (3,-1.75)
  {Active oriented links = 4 (red).\\
   Again $\|T_j|\psi_0\rangle\|^2=4$ for this $j$.};
\end{scope}

% Legend
\begin{scope}[xshift=0.0cm,yshift=-7.2cm]
  \draw[allhop] (0,0) -- (1.1,0);
  \node[lab,anchor=west] at (1.25,0) {all oriented terms in $T_j$ (incl.\ h.c.)};
  \draw[active] (0,-0.7) -- (1.1,-0.7);
  \node[lab,anchor=west] at (1.25,-0.7) {active for the chosen $D/0$ seed};
\end{scope}

\end{tikzpicture}

\caption{Active-link counting for a doublon/empty seed. }
\label{doublonlink}
\end{figure}

In the Figure \eqref{doublonlink} we can see this scenario more clearly.
Each rung has two orbitals
\((\alpha_j,\beta_j)\). A doublon, as described before, is \(D\equiv(n_j^\alpha,n_j^\beta)=(1,1)\) (two black dots on the
rung) and an empty site is \(0\equiv(0,0)\) (two white dots). The local operator \(T_j\) in
\(H_\Delta=2\Delta\sum_j T_j\) acts only on the same-parity pair \((j-1,j+1)\) and contains four
hopping channels (and their hermitian conjugates) between these two rungs. 

For an active pair
\(D\!-\!0\) (top panel) or \(0\!-\!D\) (bottom panel), exactly four oriented hops are Pauli-allowed, shown using red arrows,
creating four mutually orthogonal two-site Fock states with unit coefficients (up to signs), so
\(\|T_j|\psi_0\rangle\|^2=4\) when the link number \(\chi_j=1\). For inactive pairs \(D\!-\!D\) or \(0\!-\!0\), all hops are
blocked and \(T_j|\psi_0\rangle=0\). Summing over \(j\) for the doublon/empty seed case then gives
\(\|H_\Delta|\psi_0\rangle\|^2=(2\Delta)^2\sum_j 4\chi_j=16\Delta^2 A\), hence \(b_1=4|\Delta|\sqrt{A}\)
with \(A=\sum_j\chi_j\).

\section{Algorithm for the numerics}\label{numerical_recipe}
For a ladder of $L$ rungs, the Hamiltonian contains two CLS modes per rung, $\alpha_j$ and $\beta_j$, so the total number of single-particle modes is $2L$. In the quench studied, for example, the $\beta$ - filled state the initial state is
\begin{equation}
|\psi_0\rangle=\prod_{j=1}^{L}\beta_j^\dagger |0\rangle,
\end{equation}
so that the total particle number is fixed to \(N=L\). Since the Hamiltonian preserves total fermion number, the dynamics remains entirely within the fixed-\(N\) sector.

A further exact simplification follows from the structure of the hopping term, which only connects \(j-1\) to \(j+1\). As a result, odd and even rung parities do not mix dynamically, and the Hamiltonian decomposes as
\begin{equation}
H = H_{\rm odd}+H_{\rm even}, \qquad [H_{\rm odd},H_{\rm even}]=0.
\end{equation}
For $L=16$ rungs, i.e. $2L=32$ original ladder sites, each rung-parity sector contains $8$ rungs. Since each rung carries one $\alpha$ and one $\beta$ mode, a parity sector contains $16$ CLS modes. The all-$\beta$ initial state carries $8$ particles in each parity sector, so the many-body Hilbert space of one sector has dimension
\begin{equation}
\dim \mathcal H_{\rm sec}=\binom{16}{8}=12870.
\end{equation}

The numerical procedure is then as follows. First, one constructs the exact many-body Hamiltonian \(H_{\rm sec}\) in a single parity sector, restricted to the fixed-particle-number subspace with \(N_{\rm sec}=8\). For product states whose odd and even particle numbers differ from the all-$\beta$ example, the same construction is used with the corresponding conserved pair $(N_o,N_e)$. The basis states are represented as integer-encoded bitstrings, and the action of fermionic bilinears is implemented with the appropriate fermionic sign factors. This yields a sparse Hermitian matrix \(H_{\rm sec}\) of dimension \(12870\times 12870\).

Next, starting from the sector initial state
\begin{equation}
|\psi_{0,\rm sec}\rangle = \prod_{j\in {\rm sector}} \beta_j^\dagger |0\rangle,
\end{equation}
a Hermitian Lanczos recursion is carried out. This generates an orthonormal Krylov basis
\begin{equation}
|K_0\rangle, |K_1\rangle, |K_2\rangle, \dots
\end{equation}
with
\begin{equation}
|K_0\rangle = |\psi_{0,\rm sec}\rangle,
\end{equation}
and a tridiagonal Krylov Hamiltonian
\begin{equation}
T_{\rm sec}=
\begin{pmatrix}
a_0 & b_1 & 0 & \cdots \\
b_1 & a_1 & b_2 & \cdots \\
0 & b_2 & a_2 & \cdots \\
\vdots & \vdots & \vdots & \ddots
\end{pmatrix},
\end{equation}
where \(a_n\in \mathbb R\) and \(b_n\ge 0\). The sector Krylov complexity is then obtained from the time-evolved state in Krylov space,
\begin{equation}
|\phi_{\rm sec}(t)\rangle = e^{-iT_{\rm sec}t}|e_1\rangle,
\end{equation}
where \(|e_1\rangle\) denotes the first basis vector of the Krylov chain. Writing
\begin{equation}
\phi_n(t)=\langle K_n|\phi_{\rm sec}(t)\rangle,
\end{equation}
the complexity is
\begin{equation}
K_{\rm sec}(t)=\sum_{n\ge 0} n\, |\phi_n(t)|^2.
\end{equation}
To reconstruct the full-ladder dynamics, one does not return to the exponentially large many-body Hilbert space. Instead, one uses the exact tensor-product structure induced by the parity decomposition. Since the two parity sectors are identical and commuting, the full reduced dynamics is generated by the action
\begin{equation}
T_{\rm full}^{\rm(red)} = T_{\rm sec}\otimes I + I\otimes T_{\rm sec}.
\end{equation}
Rather than materializing this matrix explicitly, its action is implemented on matrices \(X\) as
\begin{equation}
X \mapsto T_{\rm sec}X + XT_{\rm sec}.
\end{equation}
A second Lanczos recursion is then performed in this reduced matrix space, starting from the factorized initial vector corresponding to the full-chain initial state. This produces a second tridiagonal matrix \(T_{\rm full}\), which is the Krylov Hamiltonian for the full ladder.

Finally, the full-ladder Krylov complexity is computed from
\begin{equation}
|\phi_{\rm full}(t)\rangle = e^{-iT_{\rm full}t}|e_1\rangle,
\end{equation}
via
\begin{equation}
K_{\rm full}(t)=\sum_{n\ge 0} n\, |\phi_n(t)|^2.
\end{equation}
Convergence is checked by increasing the Lanczos depths until the resulting \(K(t)\) curves overlap over the desired time window.
%\section{Analytic early-time Krylov growth in a Carroll scalar theory}
%\label{app:carroll_scalar}

\bibliography{main}

\end{document}